**The mouth speaks as much as the eyes: Free-ranging dogs depend on inner facial features for human recognition**

Rohan Sarkar[1][1], Tuhin Subhra Pal[1][2], Sandip Murmu[1][3], Anindita Bhadra[1]*

[1] Department of Biological Sciences, Behaviour and Ecology Lab, Indian Institute of Science Education and Research Kolkata, Kolkata, West Bengal, India, 741246

**Abstract**

The human face is a multi-signal system continuously transmitting information of identity and emotion. In shared human-animal environments, the face becomes a reliable tool of heterospecific recognition. Because humans display mixed behaviour and pose differential risk to animals, adaptable decision-making based on recognition and classification of humans confer a fitness benefit. The human-dog dyad is an ideal model to study heterospecific recognition due to their shared history, niche overlap, and cognitive co-evolution. Multiple studies on pet dogs have examined their human facial information processing. However, no study has examined these perceptual abilities in free-living populations in their natural habitat where the human-dog relationship is more complex and impacts survival. Comprehensive behavioural analysis of 416 free-ranging dogs in an approach-based task with differential facial occlusion of a human, demonstrated that these dogs recognize and discriminate between familiar and unfamiliar people. Negative behaviours like aggression and avoidance were unlikely to be displayed. Inner facial components like eyes, nose and mouth were more important than outer components like hair in human recognition. Unlike in pet dogs, the occlusion of even a single inner component of the face prevented recognition by facial cue alone. Personality and habitat conditions influenced the behavioural strategy adopted by the dogs too. Considering the ambiguous nature of human interactions, recognition and response in free-ranging dogs relied on dual assessment of identity and intent of a human based, in part, on their ontogeny. Such a cue-processing system highlights the selection pressures inherent in the unpredictable environment of free-living populations.

**Introduction**

Recognition involves the production and propagation of signals on one end and perception and identification of those signals on the receiver's end [1]. Individual recognition requires that certain signals are associated with specific individuals and response is tailored according to those signals. Recognition would be expected when animals interact repeatedly over time with multiple individuals with differing intentions. It works along a continuum such that individuals or groups of individuals are categorized in classes or social categories (rivals, mates, kin, etc.), and may be treated differently based on such categorization [2], [3]. Animals may use different cues, such as olfactory [4], acoustic [5], tactile[6], and visual[7] to recognise one another. For social living animals, including humans, recognising, and/or discriminating individuals of their own and other species, especially those who are part of its social context are advantageous and an important socio-cognitive ability[8]. Heterospecific recognition is also ecologically advantageous to navigate complex and heterogeneous environments consisting of predation risk [9], dietary [10], habitat needs[11] and competition [12]. For example, male black caps can recognize individual heterospecific rivals, like garden warblers, for up to eight months based on their songs and plumage [13].

Faces are one of the most important visual configurations for recognition as they provide us with a wealth of information about an individual's identity, age, gender, and emotion. They are an important visual tool for all major taxa in perception, communication, and recognition [14]. Non-human primate s[15], other mammal s[16], birds [17], and even invertebrates like wasps [18], [19] and crayfish [20] use facial features to identify conspecifics, nest mates and opponents respectively. Face discrimination is more efficient and specific as compared to

non- face object discrimination [21]. In the case of animals living in anthropogenic environments, like domestic animals, laboratory and zoo animals and urban adaptors, where human contact is inevitable, it is ecologically advantageous to have the abilities of heterospecific recognition of humans. Indeed, many animals have shown the ability of perceptual categorization and recognition of human individuals using facial cues. Wild American crows can learn to distinguish between threatening and non-threatening human individuals based on their facial cues [22]. In primates, the capacity to recognise heterospecific faces seems likely to be shaped by experience, in part. Chimpanzees and rhesus macaques raised and trained in the human environment seem to be better at discriminating unknown human faces than unknown conspecifics [23], [24]. Sheep are also known to have advanced face-recognition abilities and can discriminate between familiar and unfamiliar human faces from pictures [25].

Among domesticated animals, dogs share the closest association with humans and given their long history of domestication and co-evolution with humans, they have developed the ability to process information from human faces [26]. Pet dogs can differentiate human faces from those of other species [27], are able to recognize their owner's face from photographs [28], can differentiate their owner's face from that of a stranger's [29], and can discriminate between familiar persons based on their faces[30]. Recognition and differentiation prove difficult if global features (head shape and hair) are hidden and if faces are viewed in suboptimal conditions of visibility [30], [31]. Furthermore, the eye region seems to play an important role in human face processing and visual processing of human faces seems to be configural rather than part-based [32].

But so far, all research on face processing and recognition has been carried out on pet dogs, that form only 17-24% of the entire dog population [33].The range of the relationship between pet dogs and humans, on the whole is limited and for the most part in a positive context [34]. Additionally, although most pet dogs and their owners share a strong bond, the number of people a pet dog has to interact with on a daily basis is small. Free-ranging dogs, on the other hand, represent a model population that has a much more varied and multi-dimensional interaction with humans in a free-living, urban setting. In crowded areas like market places and railway stations, a free-ranging dog has to navigate the presence, and thus potential interactions, both direct and indirect, with up to 95 humans in a minute [35]. For these dogs, humans are both a source of threat [36] and resources [37]. The mixed relationship shared between these dogs and humans are also reflected in ethnographic accounts of several indigenous tribes where people and animals share a fluid and utilitarian bond. For example, in the Nuaulu tribe, humans and dogs hunt cooperatively and share the spoils but underperforming and injured dogs are left to die of neglect or starvation [38]. Interestingly, the mixed relationship between humans and free-ranging dogs can be thought of as a representative of a transitory stage in the co-evolution timeline between the stages of wild proto-dogs commensally scavenging human refuse and human-dependent pets. Understanding this relationship would provide deeper insights into the development of cooperative mechanism between dogs and humans in the co-evolution process and a non-Western perspective on human-dog relationship in developing countries

In general, free-ranging dogs are known to avoid unfamiliar humans but can form affiliative bonds on repeated positive interactions [39]. Humans play a central role in the social interaction networks of free-ranging dogs [40]. These dogs display higher direct interspecific behavioural interactions towards humans than intraspecific interactions and their sociability is correlated with human flux in a given area [35], [40]. These studies also raise the intriguing possibility of between-individual variation in response and decision-making. Consistent individual differences across time and contexts, termed personality or behavioural syndrome, has been evidenced in working and shelter dogs [41], [42]. Life experiences and environment are known to shape individual dog temperament and the varied experience free-ranging dogs have with humans provide the ideal system to study the interplay of personality and dog-human interactions [43]. Yet, no studies have been carried out to understand how these free-ranging dogs recognise, differentiate, and respond to different classes of humans, if at all. Moreover, while studies on pet dogs can be carried out in controlled, laboratory-based settings, the effect of living in a human-supervised and controlled environment would exert different socio-cognitive effects on the development of their perceptual and recognition abilities and cannot be generalised to free-ranging populations.

Furthermore, recognition studies on pet dogs had one or several of these methodological limitations: (a) the initial conditioned discrimination training phase in some of the studies might have caused the dogs to discriminate between owners' and strangers' face based on other perceptual elements, rather than true recognition; (b) use of images instead of live stimuli [30], [32]. 2-D images are deficient in the richness of features, like depth, perspective and motion and non-trained animals face difficulties in transference of recognition between images and live stimuli. Response or lack thereof from pictorial data also carries the ambiguity of whether the action of the animal is because of true recognition or inability to understand the stimuli [44]; (c) use of viewing time differences as evidence of recognition is not unanimously accepted [31]. Studies which tried to address either or both of these lacunae had pre-trial phases that rewarded the dog (through greetings) on approaching the owner [28], [31]. This may have introduced a positive bias in the subsequent choice test. Additionally, all the experiments were carried out in a laboratory set-up, that a dog may not usually face in its day-to-day life. One downside of carrying out experiments in laboratories is the type and the structure of information available to animals. While the difference in the usage and processing of cues and the subsequent response of the animal between laboratory and wild or free-living conditions has been well documented in spatial cognition and social learning tasks, we do not know the influence of the properties of the test environment in recognition tasks [45]. Moreover, free-living animals have to navigate through the noise of confounding signals inherent in their environment and the humans surrounding them, unlike a laboratory where there are no such distractions.

In the present study, we tried to address this knowledge gap in heterospecific recognition of humans by canids by using free-ranging dogs as our model organism and an experimental in-situ approach. This allowed us to capture the response and behaviour of dogs in their natural environment. We tried to improve on the limitations of the previous experiments by recording the spontaneous response of non-trained, free-ranging dogs to the presence of an experimenter with differential facial visibility. This was achieved by covering different parts of the experimenter's face using everyday clothing items (Fig 1). The face of a human provides the largest collection of cues for a dog to respond to on a "short notice". We tested two sets of dogs- one to whom the experimenter was familiar and another to whom he was unfamiliar. Approach in a trial was taken as a behavioural indication of confirmed recognition. The behaviours displayed in response to the human presence were recorded and scored (See Behaviours and Analysis for details). Since free-ranging dogs do not interact with humans the same way as pet dogs, nor are they trained to do so, the dog might not know it has to approach the human. In that case, two vocalization cues were provided after a certain amount of time to draw the attention of the dog (See Methods for details). The objectives of the paper are to examine the heterospecific recognition abilities of free-ranging dogs, if any, the role of internal and external facial features in such abilities, and their subsequent response. We hypothesize that 1) free-ranging dogs can recognize and discriminate between familiar and unfamiliar human 2) the discrimination would be expressed through their response towards said human 3) internal facial features would be more important to recognition.

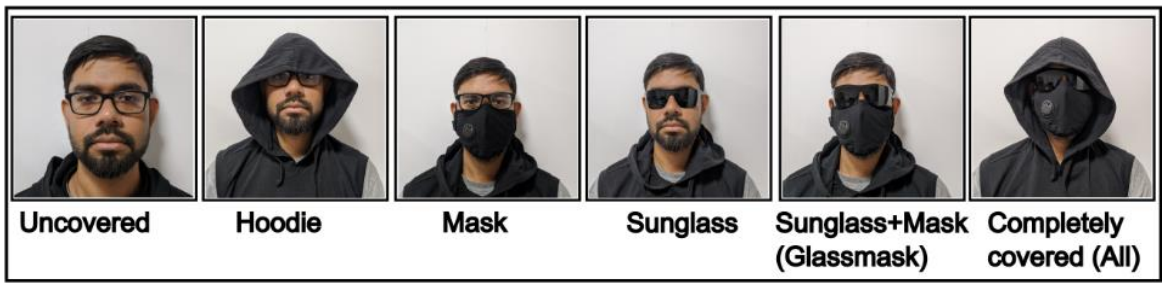

*Fig 1: Different facial cue conditions as presented to dogs*

## Results

### Q1. Do dogs recognise and distinguish between familiar and unfamiliar persons?

We analysed whether dogs showed difference in recognition, quantified through approach, between familiar and unfamiliar persons. We only considered the responses of dogs in the "Uncovered face" condition as this provided both groups of dogs with the maximum possible information in terms of facial cues. We had 43 dogs in the familiar group and 60 dogs in the unfamiliar group. The response variable, "approach", was binary (yes/no). We ran a hierarchical logistic regression with "random effect" of individual dog identity ("dogid") nested within "place". The predictor, "group" explained the relationship of experimenter to the dogs and had two levels, familiar and unfamiliar. The AIC value was 80.21.

approach ~ 1 + group + (1 | place/dogid)

The model performed well on class separability (AUROC = 0.99) and prediction accuracy (85.71%). The model-estimated marginal means of the probability of approach for familiar and unfamiliar dogs were 0.977 (95% CI: 0.827, 0.997) and 0.217 (95% CI: 0.110, 0.383) respectively (Fig 2). The results of the regression showed that familiar dogs were more likely to approach the experimenter than unfamiliar dogs (familiar - unfamiliar: 5.02 (0.76 in response scale); $p < 0.001$). The fixed effects (marginal $R^2$) account for 65.3% of the variance.

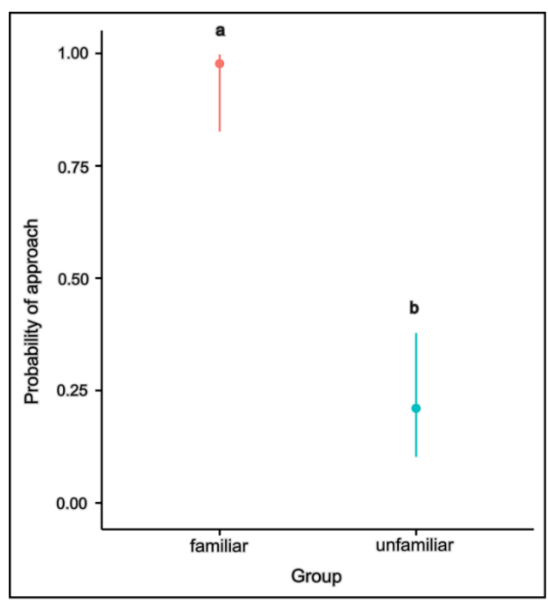

*Fig 2: Model-estimated marginal means of approaching probability of a dog when experimenter is familiar versus unfamiliar*

**Q2. Do dogs approach familiar person quicker than unfamiliar person?**

We analysed whether there is a difference in latency of approach between dogs in unfamiliar and familiar group. We defined the time from when the experimenter took position to the time when the dog approached as the approach latency. Higher the approach latency, longer the dog has taken to approach. The maximum time taken by a dog to approach the experimenter was 33s which was taken as the censored threshold value. The outcome variable is "latency", the time taken to approach the experimenter and the predictor, "group" denotes the relationship of the experimenter to the dog, "familiar" and "unfamiliar". The model was run only for the "Uncovered" cue.

$$(\text{latency, censored}) \sim \text{group}$$

The results showed that the median approach latency for the familiar group is 5s (survival/ no approach probability = 0.5). Since the unfamiliar group had an approach (event) probability of < 0.25 even at the end of the test, the median could not be computed (Fig.S1). The log-rank test gives a p-value < 0.05 and chi-square statistic of 114 indicating that the familiar and unfamiliar group differ significantly in approach latency. The RMST curve showed us that across 16.5 seconds of trial (half the time of total time, 33s, of trial), dogs in the familiar group took 6.93s to approach on average. The corresponding RMST for unfamiliar group was 16.20s. The difference of 9.27s (95% CI: -10.51, -8.02; p < 0.05) was statistically significant. Thus, dogs in the familiar group were quicker to approach than dogs in unfamiliar group (Fig 3).

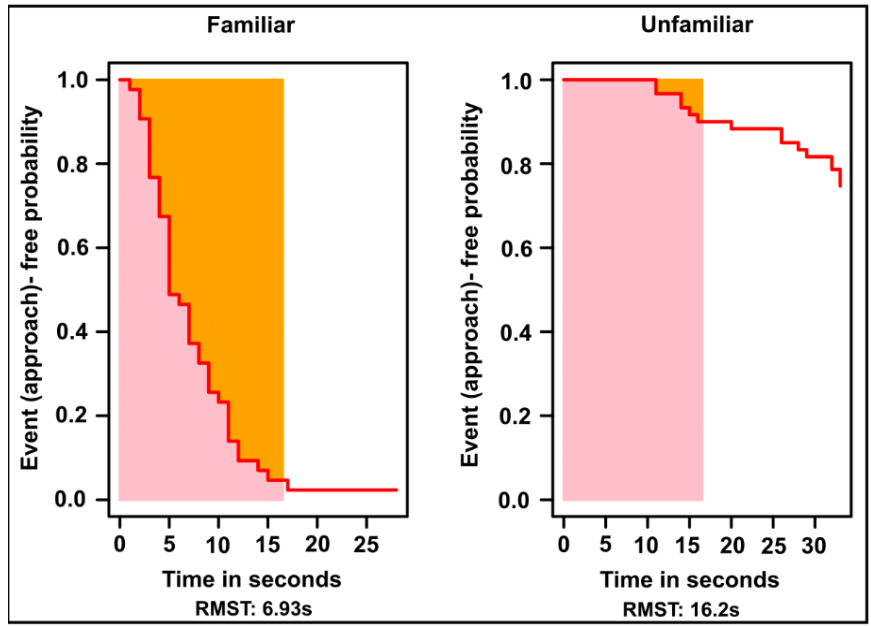

*Fig 3.* RMST as the area under the Kaplan–Meier curve up to 16.5s for familiar vs unfamiliar group

### Q3. Do dogs in the familiar group show a difference in approach when presented with different facial cues?

We analysed whether approach behaviour by dogs was affected by the amount and type of facial features visible. The proportion of approach for each of the cues are as follows: (a) uncovered: 42/43 (97.67%); (b) hoodie-covered face: 34/39(87.17%); (c) mask-covered face: 36/42(85.71%); (d) sunglass-covered face: 37/43(86.04%); (e) sunglass + mask covered face: 28/42(66.66%); (f) all-covered face: 25/39(64.10%). The response variable, "approach", was binary (yes/no). We ran a mixed effects logistic regression with "random effect" of individual dog identity ("dogid"). The predictor, "cue" refers to the six facial cues presented to the dogs.

approach ~ 1 + cue + (1 | dogid)

Furthermore, we carried out pairwise comparisons between all the levels of the cue predictor using the "fdr" (false discovery rate) method for adjusting p-values for multiple comparisons. We only considered those terms that showed significance after the adjustment. The AIC value was 201.7.

The model performed well on class separability (AUROC = 0.94) and prediction accuracy (93.75%). The model-estimated marginal means of the probability of approach for familiar dogs were as follows: (i) Uncovered face: 0.986; (ii) Hoodie covered face: 0.880; (iii) Mask covered face: 0.859; (iv) Sunglass covered face: 0.862; (v) Sunglass + Mask covered face: 0.596; (vi) All-covered: 0.570 (Fig. 4A). The results of the regression showed that dogs were less likely to approach the experimenter when his face and head were fully covered (all covered) or the internal features were not visible (sunglass + mask) as compared to other conditions. There was no difference in approach probability between uncovered face and the three other cues where only one of the features was hidden. The significant model statistics are provided in Table S1.

The grouping variable, "dogid" vary in conditional average log-odds of approach by about 1.86 SD (after controlling for fixed and random effects; Fig.S2). The ICC indicates that about 51.2% of the explained variance comes from the random effects in the model. The fixed effects (marginal R2) account for 23.3% of the variance and the conditional R2 is 62.6%.

**Q4. Was the facial cue enough for dogs to recognize or initiate recognition towards a familiar experimenter?**

We quantified the number of times confirmed recognition or recognition initiation in the dogs (dogs with scores of 5 and above) happened in the silent (only facial cue) phase out of all the successful recognition trials in each of the six conditions. We carried out a series of two-tailed binomial tests to test if the probability of such a recognition event happening due to facial cue was statistically different than chance. We did not consider all hidden face as it did not qualify the recognition criteria (Table S2). We found that for the conditions uncovered face (31/42; p-value = 0.0028), and hoodie covered face (26/34; p-value= 0.0029), more dogs responded to the facial cue than what would occur because of chance. For the conditions mask covered face (22/36; p-value = 0.2430), sunglass covered face (20/37; p-value = 0.7428), and sunglass+mask (16/28; p-value = 0.5716) the null hypothesis could not be rejected (Fig 4B).

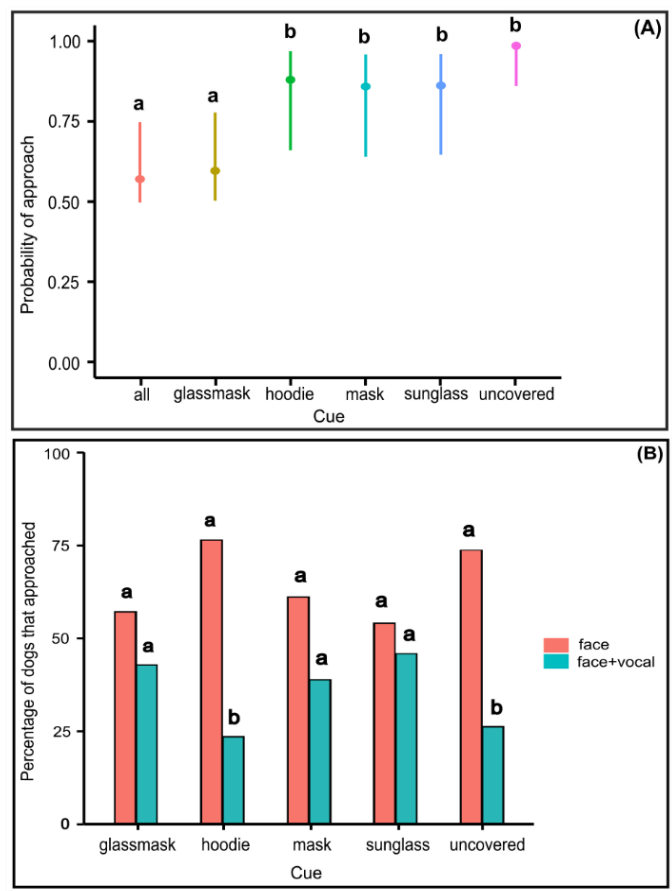

*Fig 4:* A) Predicted probability of a familiar dog approaching experimenter under different cue conditions. B) Percentage of dogs that approached on only face cue and those that required the additional vocalisation cue

**Q5. Do dogs in the unfamiliar group show a difference in approach when presented with different facial cues?**

We analysed whether approach behaviour by dogs was affected by the amount and type of facial features visible. The response variable, "approach", was binary (yes/no). We ran a mixed effects logistic regression with "random effect" of individual dog identity ("dogid"). The predictor, "cue" refers to the six facial cues presented to the dogs.

approach ~ 1 + cue + (1 | place/dogid)

There was no difference in the probability of approach behaviour in unfamiliar dogs between different facial cue conditions.

**Q6. Do dogs show a difference in behavioural scores between familiar and unfamiliar groups for each of the facial cues presented?**

We examined the behavioural scores for both familiar and unfamiliar dogs in each of the cue condition separately for the silent phase (only facial cue) and vocal phase-I. Vocal phase-II scores were not investigated because of the low number of observations in the familiar group (Control cue had 1, hoodie and mask had 9, and sunglass had 8 observations). This is because, most of the dogs responded in the vocal phase-I and a vocal phase-II was not required for them. Such low number of observations create issues of low power against anything but large effects and complete separation. In all of the cue conditions, familiar dogs had consistently higher and positive median scores and were stochastically dominant than unfamiliar dogs ($p < 0.05$). Please see Table S3 for the relevant descriptive, Wilcoxon rank sum test, and VDA statistics.

**Q7. What is the most common behaviour type shown by familiar and unfamiliar dogs when presented with each of the cues?**

We assigned a behaviour type based on the scores for each dog. The list has been provided below

| Scores | Behaviour Type |
| --- | --- |
| -5 or less | agitated |
| (-3, -4) | wary |
| (-1, -2) | alert |
| 0 | neutral |
| (1, 2) | relaxed |
| (3, 4) | affiliative |
| 5 or more | excited |

*Table 1: List of behaviour types based on scores*

We carried out an omnibus chi square goodness of fit tests, followed by Holm-corrected, pairwise comparisons in case of significance. The null hypothesis was that all behaviour types are equally common.

**Unfamiliar group**

We combined the relaxed, affiliative, and excited types into a single type, "positive" because of low frequency (sum < 5 across all cues in silent phase). The significant pairwise comparisons between behaviour types of dogs that were present during that particular phase are provided in Tables S4.A.1-S4.A.6. Dogs which had already approached were not represented for their behaviour type in the subsequent phase. Neutral behaviour type was dominant in 5/6 conditions in the silent phase. Different dogs displayed different behaviour types after the first vocalisation. No one behaviour was dominant and the differences observed were between two numerically extreme, adjacent behaviour types. Subsequent vocalisation caused dogs to display neutral behaviour type over all others in certain conditions or at least more than the extreme behaviour types (Fig 5).

**Familiar group**

We combined wary and agitated behaviour types into a single type, "high negative reaction (hnr)" because of their low frequencies of occurrence (sum <= 5 across all cues in silent and vocal phase-I). The significant pairwise comparisons between behaviour types of dogs that were present during that particular phase are provided in Tables S4.B.1-S4.B.6. Dogs which had already approached were not represented for their behaviour type in the subsequent phase. Comparisons were not carried out on vocal phase-II because of very few observations for each of the behaviour type. This could lead to erroneous results. Excited behaviour type was the dominant behaviour type in multiple conditions but as more of the internal features were covered, dogs displayed other behaviour types along with excited. The first vocalisation phase was carried out on dogs who had not approached during the silent phase and within this reduced number, dogs display excited behaviour type with uncovered face or one of the features partially covered. With complete coverage of internal features or entire head region, no one behaviour type is dominant (Fig 5).

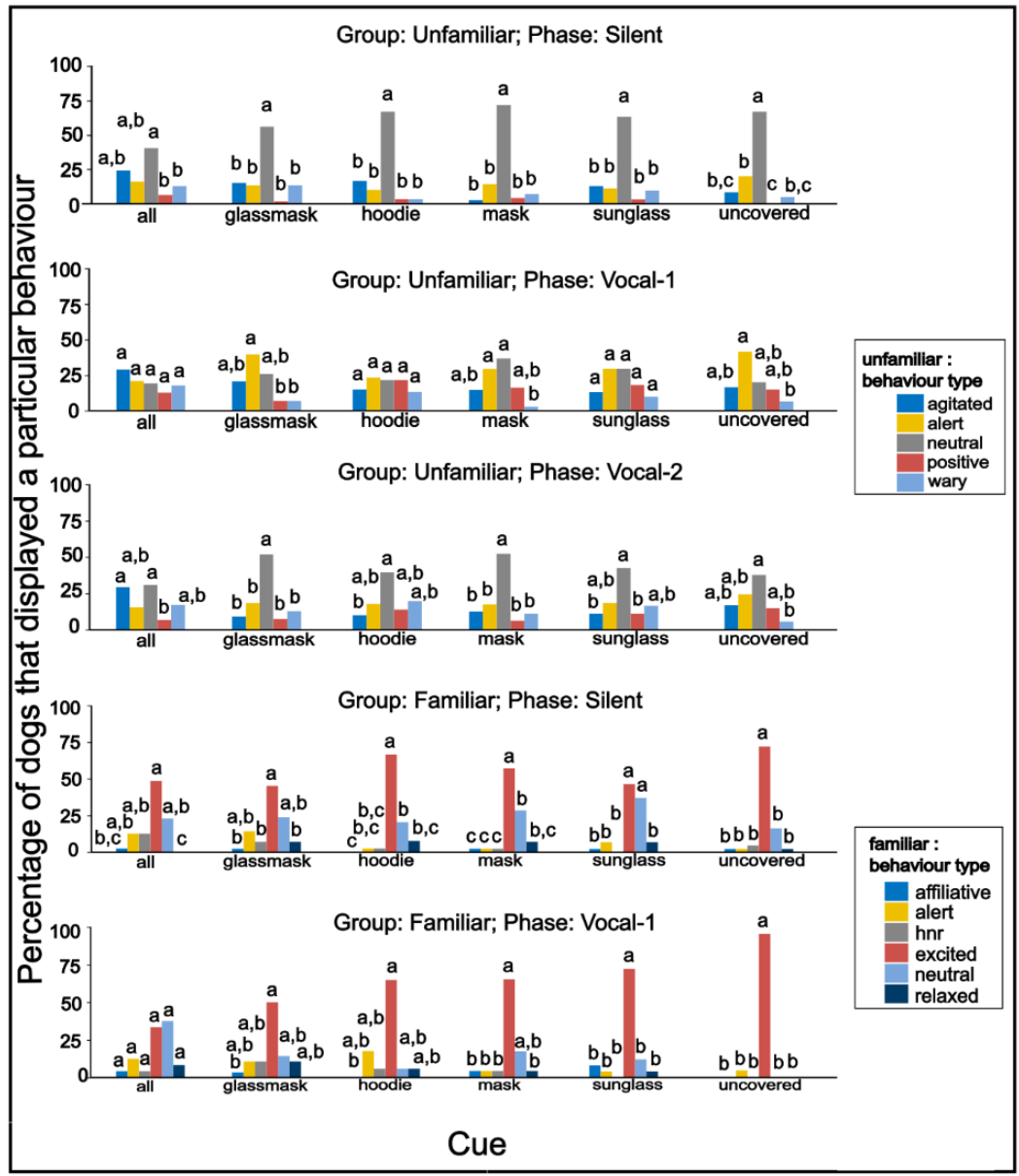

*Fig 5:* *Percentage of different behaviour types across cues and phases for unfamiliar and familiar dogs*

**Q8. Do dogs show a difference in active negative behaviours depending on the familiarity of the experimenter?**

We categorised negative behaviours on the basis of aggressiveness and active aversion as active negative behaviour. These behaviours are a subset of high negative reaction behaviour type as mentioned above and include dogs that displayed behaviours like barking, running away from experimenter, and hiding among others or a combination thereof. In general, dogs with scores of -4 and lower came under this category along with barking and growling (which have a score of -3). We did not consider dogs who displayed passive negative behaviours like pricking ears, continuous gazing, sitting up and others even if the combination of such behaviours resulted in a score of -3 or -4.

We analysed whether the display of active negative behaviours was influenced by the familiarity or unfamiliarity of the experimenter to the dogs. We compared both the groups across all the cues in the silent phase. This was because the frequency of occurrence of active negative behaviour in vocal phase-I and vocal phase-II of the familiar group were ~0 in most of the cases. The response variable was binary (whether or not active negative behaviour, anb, has been displayed). We ran a hierarchical logistic regression with "random effect" of individual dog identity ("dogid") nested within "place". The predictor, "group" explained the relationship of experimenter to the dogs and had two levels, familiar and unfamiliar. The AIC value was 485.10.

anb ~ 1 + group + (1 | place/dogid)

The model had acceptable class separability (AUROC = 0.83) and prediction accuracy (90.4%). The model-estimated marginal means for familiar and unfamiliar dogs were 0.050 (0.018, 0.124) and 0.204 (0.139, 0.288) respectively (Fig 6; Also check, Table S5). The results of the regression showed that familiar dogs were less likely to show active negative behaviour towards the experimenter than unfamiliar dogs (familiar - unfamiliar: -1.61; p = 0.0032). The places and dogs themselves vary in conditional average active negative behaviour log odds by 0.407 and 0.376 SD respectively. The fixed effects (marginal R2) account for 14.8% of the explained variance whereas 8.5% is explained by the random effects of the model.

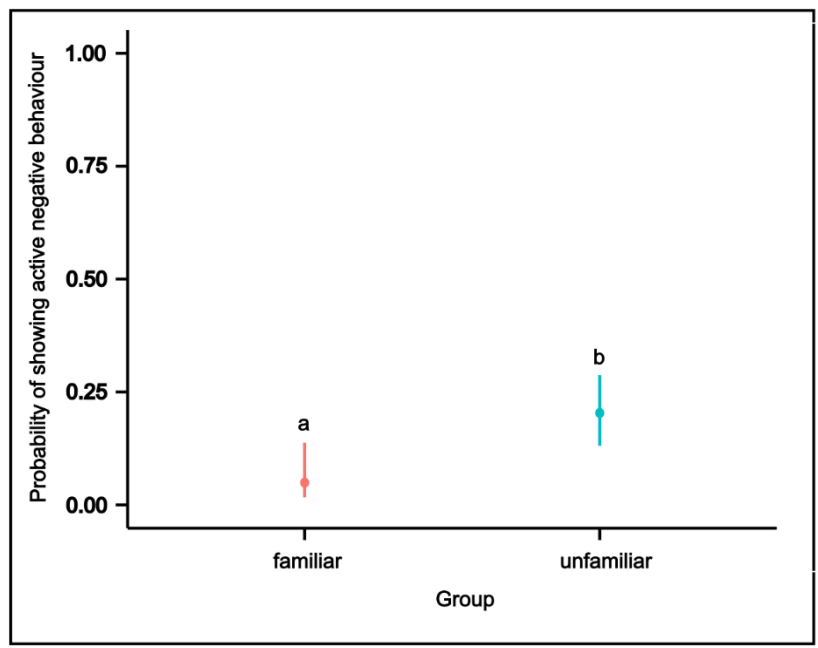

*Fig 6: Model-estimated marginal means of active negative behaviour probability in familiar versus unfamiliar group*

**Q9. Does the display of active negative behaviours vary with the type of facial cue and the phase presented to unfamiliar dogs?**

We investigated the difference in the likelihood of active negative behaviours in response to a combination of different facial cues and presence or absence of the vocalization cue in unfamiliar dogs. The response variable was binary (whether or not active negative behaviour, anb, has been displayed). We ran a hierarchical logistic regression with "random effect" of individual dog identity ("dogid") nested within "place". There was an interaction term comprised of the predictors, "cue" that consisted of all the six facial cues and, "phase" (silent, vocal phase-I, and vocal phase-II). The AIC value was 912.3.

anb ~ 1 + cue * phase + (1 | place/dogid)

The model performed well on class separability (AUROC = 0.98) and prediction accuracy (84.18%). We had a variation inflation factor (vif) of 1 between the predictors. The conditional average active negative behaviour log-odds varies between dogs by 7.97 SD (Fig.S3). The model-estimated marginal means are given in Table S6.

The results of the regression showed that active negative behaviour was less likely to occur when uncovered and mask cues were presented as compared to when facial features of the experimenter were completely covered (all-covered) or covered by sunglass and mask in the silent phase. The relevant model-statistics for the pairwise comparisons are given in the table below:

| contrast | estimate | Std. error | z-ratio | p-value |
| --- | --- | --- | --- | --- |
| Uncovered – all covered | -4.89118 | 1.81 | -2.708 | 0.0407 |
| Uncovered – sunglass & mask | -4.02455 | 1.58 | -2.548 | 0.0407 |
| All covered – mask covered | 4.42864 | 1.66 | 2.665 | 0.0407 |
| Sunglass & mask – mask covered | 3.56200 | 1.36 | 2.626 | 0.0407 |

*Table 2: Model-estimated pairwise statistics for significant contrasts*

No such differences between the cues were seen in vocal phase-I and vocal phase-II. Furthermore, active negative behaviours were more likely to occur in vocal phase-I (s – vo; est: -3.04339, p-value = 0.0306) and vocal phase-II (s - vt; est: -3.12777; p-value = 0.0306) phases of uncovered face cue as compared to its silent phase and the vocal phase-II (s – vt; est: -3.12817; p-value = 0.0053)) of mask cue over its silent phase. The fixed effects (marginal R2) account for 3.1% and the random effects account for 95.1% of the explained variance.

### Q10. Does the display of active negative behaviours vary with the type of facial cue and the phase presented to familiar dogs?

We investigated the difference in the likelihood of active negative behaviours in response to a combination of different facial cues and presence or absence of the vocalization cue (silent and vocal phase-I) in familiar dogs. The response variable was binary (whether or not active negative behaviour, anb, has been displayed). We ran a mixed effects logistic regression with "random effect" of individual dog identity ("dogid"). There was an interaction term comprised of the predictors, "cue" that consisted of all the six facial cues and, "phase" (silent and vocal phase-I).

anb ~ 1 + cue * phase + (1 | dogid)

There was no difference in the probability of active negative behaviours displayed between cues and phases in familiar dogs.

**Discussion**

Free-ranging dogs can recognize and discriminate between familiar and unfamiliar humans and such recognition is swift. In this study, dogs always responded more positively to a familiar human than an unfamiliar one, regardless of the cue presented. Inner facial elements were more important than outer elements in recognition of familiar humans. Covering the entirety of the face, including the head region, hindered recognition in general whereas wearing a sunglass or mask prevented recognition of a familiar human by facial cue alone. Dogs displayed a highly positive reaction to the presence of a familiar human, termed as "excited" behaviour type in the study, the numbers of which decreased as more of the emotional-cue producing features were hidden. The dogs were indifferent to unfamiliar humans, unless the entire head was completely covered. Vocalization from a familiar person had a positive effect on behaviour type with full or partial visibility of inner features. Initial positive vocalization by an unfamiliar human affected each dog differently with behaviour types ranging from positive to negative. Subsequent positive vocalization had a moderating effect if the dogs could see some parts of the face and head. Completely covering the head and face consistently elicited negative behaviour type with vocalization causing an increase in the intensity of such behaviour. The dogs were highly unlikely to show aggression or active avoidance behaviour on seeing a human, regardless of the cue or familiarity (Table S6). If such a negative response did occur, it was more likely to occur against an unfamiliar human than a familiar one, especially if the entirety of the head and/or face was occluded. Apart from visibility, the behavioural syndrome of individual dogs played a role in their response to the presence of a human, regardless of familiarity. Recognition and the response to it is hypothesized to be a two-step process requiring identification of the human and their intentions and is affected by the facial cue available along with the personality of individual dogs and the processes shaping it.

The dogs approached the familiar experimenter with no prior testing, training, and reward, displaying a natural, voluntary, and highly motivated response to the presence of the experimenter in the facial cue only phase (silent phase). It was obvious from the results that even partially covering one of the internal features caused difficulties in recognition by facial cue alone and required an additional vocalisation cue. In addition, the decision to approach varied between individual dogs with some dogs being cautious about approaching despite the familiarity. Two dogs never approached the experimenter except for when the entire face was visible and two more didn't approach for 4/6 conditions. On the other hand, some dogs approached as soon as the experimenter took position even if his entire face was covered. In contrary to previous research[30], [31], [46], [47], the inability to see global features like hair and head contour did not seem to hinder recognition based on facial cues alone in our study. Our findings corroborated earlier research on the importance of the eye region in human-face recognition[29], [32], although covering the nose and mouth using a mask hindered recognition capabilities just as much as wearing a sunglass. This could be because, unlike the previous recognition studies, hiding the nose and mouth together removes access to information of a larger area than when done in isolation[48], [49]. Dogs are known to discriminate and react to human facial expressions and the inner facial features mentioned convey the most relevant cues in terms of emotions[50], [51].

We hypothesize that the contrasting findings of outer versus inner features reported here could be a result of the nature of interactions between free-ranging dogs and humans. Since free-ranging dogs encounter many people on a daily basis, both familiar and unfamiliar, intent identification is as important for them as specific individual recognition. Like other urban animals, they are at the receiving end of a continuum of human actions. Humans might be indifferent to their presence, or provide food and affiliation, or also persecute them. Responding appropriately to these "mixed messages" is an important urban adaptation. Thus, effective recognition of emotional expressions, conveyed through internal features, provide a fitness benefit. Dogs to whom the experimenter was unfamiliar rarely approached, if at all, even in the uncovered face condition along with the added positive vocalization cue. One explanation for this generalised lack of approach is that under unfamiliar conditions, the perceived risk association from a human is high enough that it is safer to be cautious.

Urban animals and urban wildlife are known to make strategic behavioural decisions based on the identity of humans and their experience with them. They show boldness and affiliative behaviour towards humans with previous positive interactions but show avoidance or hostile behaviours to unfamiliar or previously hostile humans[22], [52]. While context-dependent behavioural flexibility has been demonstrated in free-ranging dogs[53], the current experiment demonstrates similar behavioural flexibility based on identity and additionally on the cue presented. The differential response to familiar and unfamiliar humans and even among individual dogs, both in the silent phase and after vocalisation, seem to strongly hint at the role of several modulating factors like different levels of anthropogenic stress, life history, and personality in shaping their behaviour. The change in behaviour types from silent to vocalization phase in the unfamiliar group can be speculated to happen because while dogs notice the experimenter in the silent phase, they are indifferent to his presence as he is just another human around the dog. The vocalization makes them aware that the behaviour and presence of the human is directed towards them. These behavioural responses are adaptive for free-ranging dogs and may help them to assess the risk involved while exploiting anthropogenic resources. Consequently, threat-specific behavioural plasticity and tolerance are beneficial and cost-effective in terms of energy expenditure. Aggression and other active negative interactions run the risk of physical injury, chronically elevated levels of stress hormone and risk of disease transmission. The results from our experiment demonstrate that free-ranging dogs may also display such tolerance as high-energy active negative behaviours were seen in only 14.49% cases across dogs in familiar and unfamiliar group in the silent phase. Unfamiliar dogs were more likely to show such behaviours than familiar dogs, although the probability of such behaviours was low for both. Additionally, the places the dogs lived in and the dogs themselves exerted a small effect on the probability of such behaviours. Subsequent results in the unfamiliar dogs show that while losing access to most of the facial cues (all-covered and sunglass+mask covered) can make active negative behaviours more likely over uncovered and mask-covered condition, there is a big effect of the dog's own characteristics on such behaviours indicating that some individuals have a propensity towards such behavioural type that might be aggravated by the lack of facial cues and unfamiliarity

The selection pressure on dogs during the domestication process led to their development of cognitive flexibility in the heterospecific social domain. Their co-evolution with humans resulted in their ability to cooperate and form bonds with humans. Recognition of humans may, thus, have been a stepping stone in the formation of such bonds. The higher social cognitive abilities of dogs in general, and free-ranging dogs specifically are probably the reason for their success in anthropogenic urban habitats. This allows them the behavioural plasticity to respond to novel ecological and social challenges with context-specific responses. Our results suggest that such plasticity may also be a result of the frequent "pre-exposure" to the human stimuli from birth and throughout the individual's life, hinting at the role of ontogeny[54], that results in their ability to decipher human features and encode the information contained within. Moreover, the role of behavioural syndrome in the differential outcome of approach and active negative behaviour provide support to the dog-domestication idea that difference in personality in the founder group of wolves and, then the proto-dogs initiated and sustained the domestication process.

While it is clear that free-ranging dogs can do class-level recognition between familiar and unfamiliar humans, it is not yet clear whether they are capable of inter-individual recognition based on a signaller's distinctive characteristics. Subsequent research can delve into the behavioural and cognitive processes underlying such discrimination, if any. We also could not control for odour cues of the experimenter. We did not observe any overt sniffing activity, although we do not rule out the possibility of the dogs using this information. If such information is indeed used, our study reveals that it is by itself insufficient information for dogs to act on and they require facial cues on top of it for recognition. Since vocalization also seems to have an additive effect where facial cue is not enough, further experiments on the cross-modality of recognition process might shed light on whether a dog's recognition system is holistic or cue-biased. An additional finding from our experiment was the fact that behavioural syndromes and habitat play a role in a dog's response to a human. It would be interesting to test the interplay of anthropogenic stress, life histories of dogs, area-specific dog-human interactions, and behavioural syndromes that shape a dog's response to human-generated cues, as suggested by this study. Finally, the findings from this experiment can have real-world impact as it can make people on the streets aware of how their clothing and behaviour might influence a dog's behaviour towards them. Understanding the trigger for such behaviours may improve human-dog social dynamics and reduce dog-human conflict.

## Methods

### Study Area

The study was conducted in urban and semi-urban habitats in and around Kalyani, India (22.9747°N, 88.4337°E). The 14 field-sites have been highlighted in the maps (Fig.S4). The study was carried out during December 2020, January- March 2021, December 2021, August- December 2022, September- November 2023, and January 2024. All the areas had significant human presence.

## Subjects

The experiment was carried out separately on two subcategories of dogs:

(i) Familiar group: This group consisted of dogs to whom the experimenter was a familiar person. Familiarity, in this case, meant that both the dogs and the human had pre-exposure to each other. This constituted provisioning of food, affiliative vocalisation and petting on the part of the experimenter on a semi-regular basis. The dogs were thus comfortable with the experimenter's presence. A total of 43 dogs residing on the campus of IISER Kolkata were selected for the experiment. One dog migrated and three others died during the course of the experiment. The experimenter had similar levels of interaction with all of them over the years.

(ii) Unfamiliar group: This group consisted of dogs to whom the experimenter was a stranger. Neither of them had pre-exposure to each other, although the dogs had exposure to humans on a daily basis and had a varied range of interactions with them. The experiment was done on a total of 373 dogs, spread across the various facial cue conditions.

## Stimuli

A single male experimenter (RS) acted as the live stimulus with different parts of his face and head covered. The dogs were thus presented with six types of facial cues: (a) Uncovered face: Both the external (hair and head shape) and internal (eyes, nose, mouth) features were visible to the dog. (b) Mask covered face: The nose and mouth of the experimenter were covered by a mask. (c) Sunglass covered face: The eye region of the experimenter was covered by sunglasses (d) Mask and sunglass covered face: The entire internal features of the face were covered. Only hair and head contour were visible. (e) Hoodie covered face: The hair and head contour were hidden by the hoodie but the internal features were visible along with the chin. (f) All hidden face: The experimenter wore hoodie, sunglass and mask hiding all parts of his face. We used clothing to replicate the above conditions in a way that mimics the conditions that the dogs are used to seeing humans in on the streets.

## Experimental Protocol

**Familiar group:** The experimenter would walk throughout the IISER-Kolkata campus, and on spotting one of the 43 selected dogs, would position himself at a distance, "x", such that 1m < x < 6m from where the dog could clearly see the experimenter. The experimenter would be in a neutral stance with hands at the side with open palms and look straight ahead with a neutral expression on the face. The experimenter stood in silence for the first 5 seconds. This was known as the silent phase. If the focal dog did not approach the experimenter within this time, the experimenter would provide a short positive vocalisation ("aye-aye" * 2) and stand in the neutral stance for 10 more seconds. This was known as the vocal phase-I. If the dog did not approach, a second short positive vocalisation and 10 more seconds would be offered to the dog to make its approach. This was known as the vocal phase-II. These three phases taken together, constituted a single trial. The trial was stopped, regardless of the dog's response after the vocal phase-II. If the dog approached the experimenter before that, the trial was stopped at whatever point the approach was made. A dog making an approach would be rewarded with a petting.

The experimenter made sure to present his face to the dog as far as possible. He was allowed to rotate on the spot, in case the dog moved but could not leave his initial position. The uncovered face condition was carried out first for all dogs. This was the control condition on account of being the most common situation encountered by free-ranging dogs on a daily basis. The other five conditions were provided to the same dogs in a random order. Till the time a dog had completed all of its 6 phases, the experimenter maintained a neutral demeanour towards them during non-experiment time too, so as to prevent bias. The total time duration required for each trial was 30s (+ 5s). A gap of 4-5 days, at the minimum, was provided to a dog between conditions so that the response of one condition was not carried over to the next. An approach was said to be made when the dog came within petting distance of the experimenter. All responses and behaviours of the dogs up to its approach (if any) during the trial were noted. All trials were video recorded.

**Unfamiliar group:** The experimenter would walk or travel through one of the 13 field sites (excluding IISER Kolkata) and on spotting a dog that was awake would take a neutral stance in front of them at a distance, "x", as mentioned above. The silent, vocal phase-I, and vocal phase-II were carried out in the same way as above. All conditions were same as above save that, unlike the familiar group, each condition was carried out on a different set of dogs to prevent habituation to the experimenter.

### Behaviours and Analysis

We noted down the behaviours and actions of the dog in each of the phase. We created an ethogram of all the behaviours that had been displayed (Table S7). We then assigned a value and direction to each of the behaviour (positive, negative, or neutral) that together constituted the score for that particular behaviour. The direction was based on their reaction to human stimuli. The value was assigned on the basis of the level of excitement or agitation displayed and the amount of bodily motion involved. For example, pricking ears, which is a sign of alertness and vigilance and involves movement of only one body part for a short burst of time is given a score of -1 whereas continuous circular and/or loose back and forth tail wagging, which is a sign of affiliation and hence a more directed action of a longer duration is scored +2. Thus, behaviours displayed by a dog in a particular phase was added up to give their final score. We only scored established behaviours on which there had been prior research either from our lab or others or behaviours whose emotional valence could be clearly assigned. Condition-wise recognition in dogs would only be stated to occur if approach happened significantly more than chance. In phases within a trial, a cue was said to initiate recognition if the presentation of cue elicited a combination of movement towards experimenter and affiliative behaviours (a combined score of 5 and above) preceding approach. Thus, each dog had a corresponding final score for each of the 3 phases (silent, vocal phase-I and vocal phase-II) unless it had already approached the experimenter in the previous phase.

All the analyses were done in R 4.3.1[55]. We carried out generalised linear mixed effects modelling using the lme4 package[56]. We reported bias adjusted, model-estimated marginal means using the emmeans package[57], [58].Pairwise comparisons p-values were adjusted using the "false discovery rate" method and estimates were reported in log-odds scale[59], [60]. Model diagnostics were carried out using the DHARMa package[61]. Random effect variables were included in the models, despite some models having low variance components to maintain fidelity between our models and the data-generating process. In such cases, the intraclass correlation coefficient (ICC) was not reported as the relevant R packages do not provide the statistic. For proportional and categorical data with multiple categories, we carried out two-tailed binomial tests and chi square goodness of fit test respectively. A two-tailed Wilcoxon rank sum test was used to compare two independent groups. In case of significance, this was followed by the Vargha and Delaney's A (VDA) effect size statistic to determine the stochastic dominance[62]. We carried out time-to-event analysis using the survival package to compare latencies through Kaplan—Meier survival (event) estimates[63]. This was followed by the logrank test to check for statistical significance. We also reported the restricted mean survival time (RMST) as our event (approach) rate was low for the unfamiliar group[64]. We used intraclass correlation estimates based on a single rating, absolute agreement, two-way random-effects model for inter-rater reliability. The estimates were found to be 0.905 (95% CI: 0.882, 0.923) for the behavioural scores and 0.997 (95% CI: 0.995, 0.998) for latency.

## Data Availability Statement

The datasets and code presented in this study can be found in online repositories. The names of the repository/repositories and accession number(s) can be found at: [OSF | Mask Project](#)

## Ethics Statement

Ethical review and approval were not required for the animal study because the experiment did not involve any invasive procedure. This experimental protocol did not need any additional clearance from the Institute ethics committee, as it did not violate any law under the Prevention of Cruelty to Animals Act 1960 of the Parliament of India.

## Author Contributions

RSa and ABha: conceptualization, project administration, funding acquisition, and writing-reviewing and editing. RSa: data curation, formal analysis, methodology, visualization, and writing original draft. ABha: resources and supervision. TsP and ABha: validation. RSa, TsP, SM: investigation. All authors contributed to the article and approved the submitted version.

## Competing Interest

The authors declare that they have no known competing financial interests or personal relationships that could have influenced the work reported in this paper.

## Funding

Rohan Sarkar and Sandip Murmu were supported by Indian Institute of Science Education and Research-Kolkata Institute fellowship. Tuhin Subhra Pal was supported by University Grants Commission fellowship. This work was supported by the Animal Behaviour Society Developing Nations Student Research Grant 2021.

## Acknowledgement

Rohan Sarkar would like to express gratitude to Animal Behaviour Society for providing a Student Research Grant that funded this project. Rohan Sarkar would also like to thank Anwesha Acharjee and Ritwik Das for their help.

**Supplementary Information**

**Result**

**Q2.** Do dogs approach familiar person quicker than unfamiliar person?

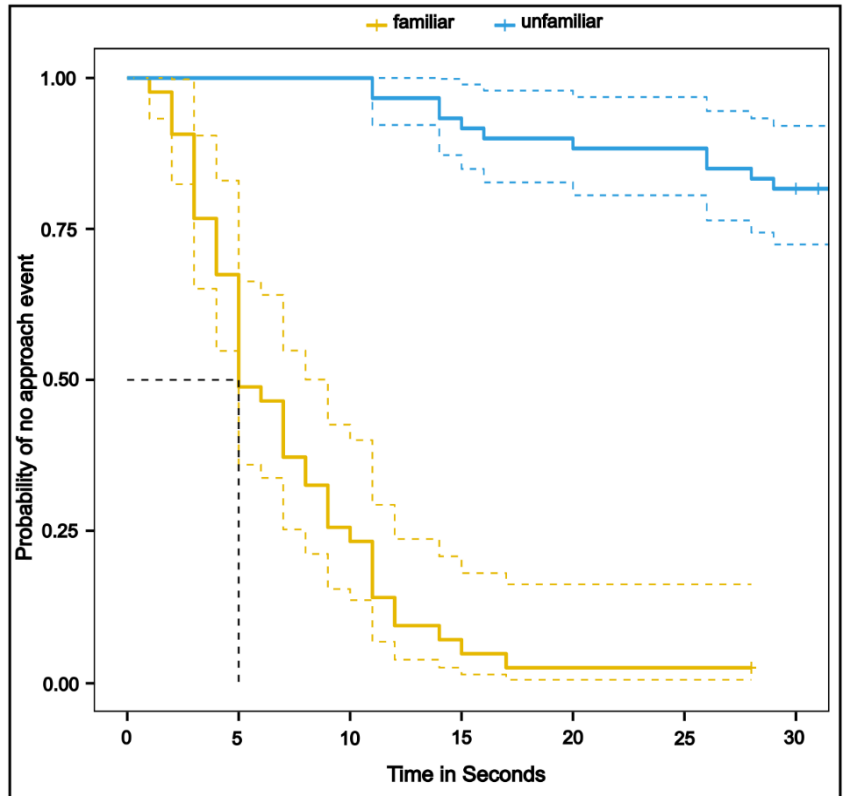

Figure S1: Kaplan-Meier plot for survival (no-approach) curves

**Q3. Do dogs in the familiar group show a difference in approach when presented with different facial cues?**

Table S1: Pairwise comparisons between the Estimated marginal means (EMMs) of all combinations between any two pairs of cues in the familiar group

| Contrast | Estimate | Std. Error | z-ratio | p-value |
|---|---|---|---|---|
| Uncovered face – All covered | 4.30 | 1.24 | 3.46 | 0.0063 |
| (Uncovered face) – (Sunglass & Mask) | 4.11 | 1.23 | 3.34 | 0.0063 |
| All covered – Hoodie covered | -2.01 | 0.74 | -2.71 | 0.0308 |
| All covered – Mask covered | -1.82 | 0.71 | -2.56 | 0.0308 |
| All covered – Sunglass covered | -1.85 | 0.70 | -2.62 | 0.0308 |
| (Sunglass & Mask) – Hoodie covered | -1.82 | 0.73 | -2.49 | 0.0318 |
| (Sunglass & Mask) – Mask covered | -1.63 | 0.69 | -2.35 | 0.0351 |
| (Sunglass & Mask) – Sunglass covered | -1.66 | 0.69 | -2.40 | 0.0346 |

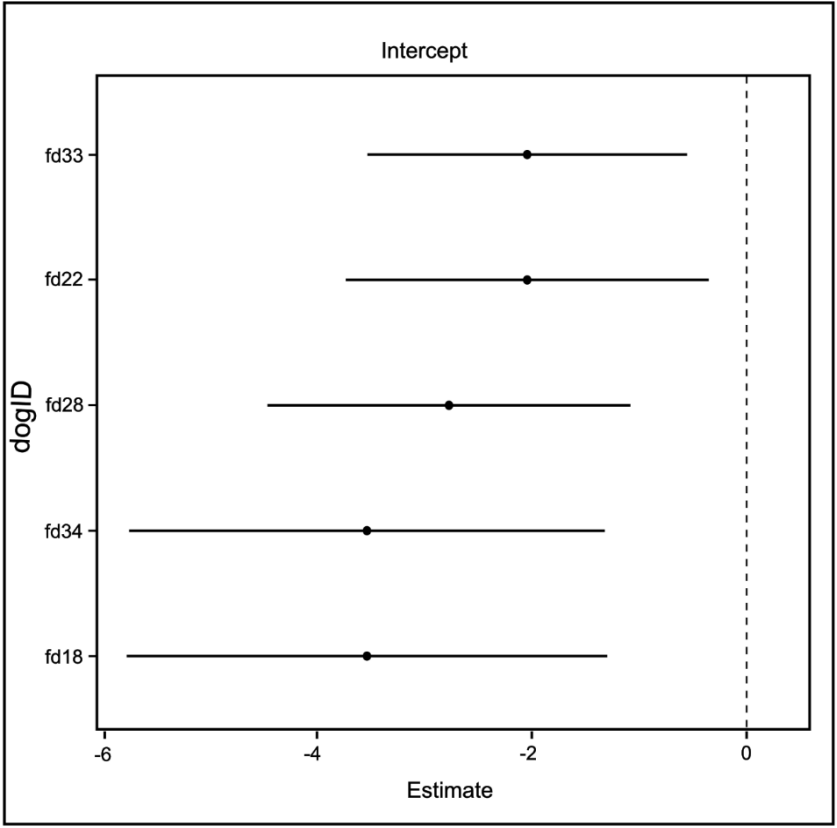

Figure S2: Random effect estimates for dogs whose 95% CI did not cross the 0 value

## Q4. Was the facial cue enough for dogs to recognize or initiate recognition towards a familiar experimenter?

Table S2: Results from the chi-square test for each cue condition to check if the number of dogs that approached the experimenter were greater than what would have approached because of chance

| Cue | n | approach | statistic | p | df | p.signif |
|---|---|---|---|---|---|---|
| Uncovered | 43 | 42 | 39.09302 | 4.04e-10 | 1 | **** |
| Hoodie | 39 | 34 | 21.5641 | 0.00000342 | 1 | **** |
| All | 39 | 25 | 3.102564 | 0.0782 | 1 | ns |
| Mask | 42 | 36 | 21.42857 | 0.00000367 | 1 | **** |
| Sunglass | 43 | 37 | 22.34884 | 0.00000227 | 1 | **** |
| Sunglass & Mask | 42 | 28 | 4.666667 | 0.0308 | 1 | * |

**Q6.** **Do dogs show a difference in behavioural scores between familiar and unfamiliar groups for each of the facial cues presented?**

Table S3: Summary and test statistics, including those of Wilcoxon rank sum test, and Vargha and Delaney's A (VDA) effect size for the difference in behavioural scores between familiar and unfamiliar dogs

| Cue | Episode | Median | Mean | W | p-value | VDA |
|---|---|---|---|---|---|---|
| Uncovered | Silent | Familiar: 7 Unfamiliar: 0 | Familiar: 5.02 Unfamiliar: -1.06 | 2277.5 | < 0.05 | 0.883 |
| Uncovered | Vocal phase-I | Familiar: 7 Unfamiliar: -1 | Familiar: 7.14 Unfamiliar: -1.35 | 1165.5 | < 0.05 | 0.963 |
| Hoodie | Silent | Familiar: 7 Unfamiliar: 0 | Familiar: 4.79 Unfamiliar: -1.21 | 2063 | < 0.05 | 0.881 |
| Hoodie | Vocal phase-I | Familiar: 7 Unfamiliar: -1 | Familiar: 4.23 Unfamiliar: -0.63 | 795 | < 0.05 | 0.903 |
| Sunglass | Silent | Familiar: 2 Unfamiliar: 0 | Familiar: 3.39 Unfamiliar: -1.17 | 2172.5 | < 0.05 | 0.815 |
| Sunglass | Vocal phase-I | Familiar: 7 Unfamiliar: -1 | Familiar: 6.12 Unfamiliar: -0.81 | 1377 | < 0.05 | 0.928 |
| Mask | Silent | Familiar: 6 Unfamiliar: 0 | Familiar: 4.21 Unfamiliar: -0.4 | 2439.5 | < 0.05 | 0.830 |
| Mask | Vocal phase-I | Familiar: 7 Unfamiliar: 0 | Familiar: 4.69 Unfamiliar: -1.01 | 1341.5 | < 0.05 | 0.895 |
| Sunglass + Mask | Silent | Familiar: 2 Unfamiliar: 0 | Familiar: 2.83 Unfamiliar: -1.64 | 1910.5 | < 0.05 | 0.771 |
| Sunglass + Mask | Vocal one | Familiar: 4 Unfamiliar: -1 | Familiar: 3.5 Unfamiliar: -1.70 | 1316.5 | < 0.05 | 0.858 |
| All covered | Silent | Familiar: 4 Unfamiliar: -2 | Familiar: 2.74 Unfamiliar: -2.51 | 1917.5 | < 0.05 | 0.793 |
| All covered | Vocal phase-I | Familiar: 0 Unfamiliar: -2 | Familiar: 2.33 Unfamiliar: -2.37 | 1220 | < 0.05 | 0.889 |

**Q7. What is the most common behaviour type shown by familiar and unfamiliar dogs when presented with each of the cue?**

Tables S4.A.1-S4.B.6: Results of the pairwise chi-square comparison tests between behaviours in each of the cue condition. Pairwise results have been presented for cues only in case the omnibus chi-square test was significant.

A) Unfamiliar

S4.A.1. Uncovered cue

| behaviour1 | behaviour2 | phase | statistic | p | df | p.adj | p.adj.signif |
|---|---|---|---|---|---|---|---|
| neutral | alert | silent | 15.07 | 1.03e-04 | 1 | 7.21e-04 | *** |
| neutral | wary | silent | 31.83 | 0.00e+00 | 1 | 2.00e-07 | **** |
| neutral | agitated | silent | 27.22 | 2.00e-07 | 1 | 1.50e-06 | **** |
| neutral | positive | silent | 40.00 | 0.00e+00 | 1 | 0.00e+00 | **** |
| alert | wary | silent | 5.40 | 2.01e-02 | 1 | 1.00e-01 | ns |
| alert | agitated | silent | 2.88 | 8.96e-02 | 1 | 2.50e-01 | ns |
| alert | positive | silent | 12.00 | 5.32e-04 | 1 | 3.19e-03 | ** |
| wary | agitated | silent | 0.50 | 4.80e-01 | 1 | 4.80e-01 | ns |
| wary | positive | silent | 3.00 | 8.33e-02 | 1 | 2.50e-01 | ns |
| agitated | positive | silent | 5.00 | 2.53e-02 | 1 | 1.01e-01 | ns |
| neutral | alert | vocal-I | 4.56 | 3.26e-02 | 1 | 0.228000 | ns |
| neutral | wary | vocal-I | 4.00 | 4.55e-02 | 1 | 0.273000 | ns |
| neutral | agitated | vocal-I | 0.18 | 6.70e-01 | 1 | 1.000000 | ns |

| behaviour1 | behaviour2 | phase | statistic | p | df | p.adj | p.adj.signif |
|---|---|---|---|---|---|---|---|
| neutral | positive | vocal-I | 0.42 | 5.13e-01 | 1 | 1.000000 | ns |
| alert | wary | vocal-I | 15.20 | 9.64e-05 | 1 | 0.000964 | *** |
| alert | agitated | vocal-I | 6.42 | 1.12e-02 | 1 | 0.089600 | ns |
| alert | positive | vocal-I | 7.52 | 6.07e-03 | 1 | 0.054600 | ns |
| wary | agitated | vocal-I | 2.57 | 1.09e-01 | 1 | 0.545000 | ns |
| wary | positive | vocal-I | 1.92 | 1.66e-01 | 1 | 0.664000 | ns |
| agitated | positive | vocal-I | 0.052 | 8.19e-01 | 1 | 1.000000 | ns |
| neutral | alert | vocal-II | 1.48 | 0.223000 | 1 | 0.89200 | ns |
| neutral | wary | vocal-II | 12.56 | 0.000393 | 1 | 0.00393 | ** |
| neutral | agitated | vocal-II | 4.17 | 0.041100 | 1 | 0.28800 | ns |
| neutral | positive | vocal-II | 5.14 | 0.023300 | 1 | 0.18600 | ns |
| alert | wary | vocal-II | 6.25 | 0.012400 | 1 | 0.11200 | ns |
| alert | agitated | vocal-II | 0.72 | 0.394000 | 1 | 0.89200 | ns |
| alert | positive | vocal-II | 1.19 | 0.275000 | 1 | 0.89200 | ns |
| wary | agitated | vocal-II | 3.00 | 0.083300 | 1 | 0.50000 | ns |
| wary | positive | vocal-II | 2.27 | 0.132000 | 1 | 0.66000 | ns |
| agitated | positive | vocal-II | 0.05 | 0.808000 | 1 | 0.89200 | ns |
| behaviour1 | behaviour2 | phase | statistic | p | df | p.adj | p.adj.signif |

S4.A.2. Hoodie-covered cue

| behaviour1 | behaviour2 | phase | statistic | p | df | p.adj | p.adj.signif |
|---|---|---|---|---|---|---|---|
| neutral | alert | silent | 25.13 | 5.00e-07 | 1 | 4.30e-06 | **** |
| neutral | wary | silent | 34.38 | 0.00e+00 | 1 | 0.00e+00 | **** |
| neutral | agitated | silent | 18.00 | 2.21e-05 | 1 | 1.55e-04 | *** |
| neutral | positive | silent | 34.38 | 0.00e+00 | 1 | 0.00e+00 | **** |
| alert | wary | silent | 2.00 | 1.57e-01 | 1 | 6.28e-01 | ns |
| alert | agitated | silent | 1.00 | 3.17e-01 | 1 | 6.34e-01 | ns |
| alert | positive | silent | 2.00 | 1.57e-01 | 1 | 6.28e-01 | ns |
| wary | agitated | silent | 5.33 | 2.09e-02 | 1 | 1.25e-01 | ns |
| wary | positive | silent | 0.00 | 1.00e+00 | 1 | 1.00e+00 | ns |
| agitated | positive | silent | 5.33 | 2.09e-02 | 1 | 1.25e-01 | ns |
| neutral | alert | vocal-II | 4.17 | 0.0411 | 1 | 0.329 | ns |
| neutral | wary | vocal-II | 3.33 | 0.0679 | 1 | 0.475 | ns |
| neutral | agitated | vocal-II | 9.00 | 0.0027 | 1 | 0.027 | * |
| neutral | positive | vocal-II | 6.25 | 0.0124 | 1 | 0.112 | ns |
| alert | wary | vocal-II | 0.05 | 0.8190 | 1 | 1.000 | ns |
| alert | agitated | vocal-II | 1.14 | 0.2850 | 1 | 1.000 | ns |

| behaviour1 | behaviour2 | phase | statistic | p | df | p.adj | p.adj.signif |
|---|---|---|---|---|---|---|---|
| alert | positive | vocal-II | 0.25 | 0.6170 | 1 | 1.000 | ns |
| wary | agitated | vocal-II | 1.66 | 0.1970 | 1 | 1.000 | ns |
| wary | positive | vocal-II | 0.52 | 0.4670 | 1 | 1.000 | ns |
| agitated | positive | vocal-II | 0.33 | 0.5640 | 1 | 1.000 | ns |

S4.A.3. All-covered cue

| behaviour1 | behaviour2 | phase | statistic | p | df | p.adj | p.adj.signif |
|---|---|---|---|---|---|---|---|
| neutral | alert | silent | 6.42 | 1.12e-02 | 1 | 0.089600 | ns |
| neutral | wary | silent | 8.75 | 3.08e-03 | 1 | 0.027700 | * |
| neutral | agitated | silent | 2.50 | 1.14e-01 | 1 | 0.654000 | ns |
| neutral | positive | silent | 15.20 | 9.64e-05 | 1 | 0.000964 | *** |
| alert | wary | silent | 0.22 | 6.37e-01 | 1 | 0.744000 | ns |
| alert | agitated | silent | 1.00 | 3.17e-01 | 1 | 0.744000 | ns |
| alert | positive | silent | 2.57 | 1.09e-01 | 1 | 0.654000 | ns |
| wary | agitated | silent | 2.13 | 1.44e-01 | 1 | 0.654000 | ns |
| wary | positive | silent | 1.33 | 2.48e-01 | 1 | 0.744000 | ns |
| agitated | positive | silent | 6.36 | 1.16e-02 | 1 | 0.089600 | ns |
| neutral | alert | vocal-II | 3.00 | 0.08330 | 1 | 0.6660 | ns |
| neutral | wary | vocal-II | 2.28 | 0.13100 | 1 | 0.7630 | ns |
| neutral | agitated | vocal-II | 0.02 | 0.86600 | 1 | 1.0000 | ns |
| neutral | positive | vocal-II | 8.90 | 0.00284 | 1 | 0.0284 | * |
| alert | wary | vocal-II | 0.05 | 0.81900 | 1 | 1.0000 | ns |
| alert | agitated | vocal-II | 2.46 | 0.11700 | 1 | 0.7630 | ns |

S4.A. All-covered cue

| behaviour1 | behaviour2 | phase | statistic | p | df | p.adj | p.adj.signif |
|---|---|---|---|---|---|---|---|
| alert | positive | vocal-II | 1.92 | 0.16600 | 1 | 0.7630 | ns |
| wary | agitated | vocal-II | 1.81 | 0.17800 | 1 | 0.7630 | ns |
| wary | positive | vocal-II | 2.57 | 0.10900 | 1 | 0.7630 | ns |
| agitated | positive | vocal-II | 8.04 | 0.00456 | 1 | 0.0410 | * |

S4.A.4. Mask-covered cue

| behaviour1 | behaviour2 | phase | statistic | p | df | p.adj | p.adj.signif |
|---|---|---|---|---|---|---|---|
| neutral | alert | silent | 26.66 | 2.00e-07 | 1 | 1.70e-06 | **** |
| neutral | wary | silent | 36.81 | 0.00e+00 | 1 | 0.00e+00 | **** |
| neutral | agitated | silent | 44.30 | 0.00e+00 | 1 | 0.00e+00 | **** |
| neutral | positive | silent | 41.67 | 0.00e+00 | 1 | 0.00e+00 | **** |
| alert | wary | silent | 1.66 | 1.97e-01 | 1 | 7.88e-01 | ns |
| alert | agitated | silent | 5.33 | 2.09e-02 | 1 | 1.25e-01 | ns |
| alert | positive | silent | 3.76 | 5.22e-02 | 1 | 2.61e-01 | ns |
| wary | agitated | silent | 1.28 | 2.57e-01 | 1 | 7.88e-01 | ns |
| wary | positive | silent | 0.50 | 4.80e-01 | 1 | 9.60e-01 | ns |
| agitated | positive | silent | 0.20 | 6.55e-01 | 1 | 9.60e-01 | ns |
| neutral | alert | vocal-I | 0.55 | 4.56e-01 | 1 | 9.12e-01 | ns |
| neutral | wary | vocal-I | 19.59 | 9.60e-06 | 1 | 9.58e-05 | **** |
| neutral | agitated | vocal-I | 6.42 | 1.12e-02 | 1 | 8.96e-02 | ns |
| neutral | positive | vocal-I | 5.44 | 1.96e-02 | 1 | 1.18e-01 | ns |
| alert | wary | vocal-I | 14.72 | 1.24e-04 | 1 | 1.12e-03 | ** |
| alert | agitated | vocal-I | 3.33 | 6.79e-02 | 1 | 2.72e-01 | ns |
| alert | positive | vocal-I | 2.61 | 1.06e-01 | 1 | 3.18e-01 | ns |
| wary | agitated | vocal-I | 5.33 | 2.09e-02 | 1 | 1.18e-01 | ns |
| wary | positive | vocal-I | 6.23 | 1.26e-02 | 1 | 8.96e-02 | ns |
| agitated | positive | vocal-I | 0.04 | 8.27e-01 | 1 | 9.12e-01 | ns |
| neutral | alert | vocal-II | 11.00 | 9.11e-04 | 1 | 6.38e-03 | ** |
| neutral | wary | vocal-II | 16.90 | 3.94e-05 | 1 | 3.55e-04 | *** |
| neutral | agitated | vocal-II | 15.24 | 9.45e-05 | 1 | 7.56e-04 | *** |
| neutral | positive | vocal-II | 22.72 | 1.90e-06 | 1 | 1.86e-05 | **** |
| alert | wary | vocal-II | 0.88 | 3.46e-01 | 1 | 1.00e+00 | ns |
| alert | agitated | vocal-II | 0.47 | 4.91e-01 | 1 | 1.00e+00 | ns |
| alert | positive | vocal-II | 3.26 | 7.07e-02 | 1 | 4.24e-01 | ns |
| wary | agitated | vocal-II | 0.06 | 7.96e-01 | 1 | 1.00e+00 | ns |
| wary | positive | vocal-II | 0.81 | 3.66e-01 | 1 | 1.00e+00 | ns |
| agitated | positive | vocal-II | 1.33 | 2.48e-01 | 1 | 1.00e+00 | ns |

S4.A.5 Sunglass-covered cue

| behaviour1 | behaviour2 | phase | statistic | p | df | p.adj | p.adj.signif |
|---|---|---|---|---|---|---|---|
| neutral | alert | silent | 22.26 | 2.40e-06 | 1 | 1.90e-05 | **** |
| neutral | wary | silent | 24.20 | 9.00e-07 | 1 | 7.80e-06 | **** |
| neutral | agitated | silent | 20.44 | 6.10e-06 | 1 | 4.29e-05 | **** |
| neutral | positive | silent | 33.39 | 0.00e+00 | 1 | 1.00e-07 | **** |
| alert | wary | silent | 0.07 | 7.82e-01 | 1 | 1.00e+00 | ns |
| alert | agitated | silent | 0.06 | 7.96e-01 | 1 | 1.00e+00 | ns |
| alert | positive | silent | 2.77 | 9.56e-02 | 1 | 4.78e-01 | ns |
| wary | agitated | silent | 0.28 | 5.93e-01 | 1 | 1.00e+00 | ns |
| wary | positive | silent | 2.00 | 1.57e-01 | 1 | 6.28e-01 | ns |
| agitated | positive | silent | 3.60 | 5.78e-02 | 1 | 3.47e-01 | ns |
| neutral | alert | vocal-I | 0.00 | 1.0000 | 1 | 1.000 | ns |
| neutral | wary | vocal-I | 6.00 | 0.0143 | 1 | 0.143 | ns |
| neutral | agitated | vocal-I | 3.84 | 0.0499 | 1 | 0.399 | ns |
| neutral | positive | vocal-I | 1.68 | 0.1940 | 1 | 1.000 | ns |
| alert | wary | vocal-I | 6.00 | 0.0143 | 1 | 0.143 | ns |
| alert | agitated | vocal-I | 3.84 | 0.0499 | 1 | 0.399 | ns |

| behaviour1 | behaviour2 | phase | statistic | p | df | p.adj | p.adj.signif |
|---|---|---|---|---|---|---|---|
| alert | positive | vocal-I | 1.68 | 0.1940 | 1 | 1.000 | ns |
| wary | agitated | vocal-I | 0.28 | 0.5930 | 1 | 1.000 | ns |
| wary | positive | vocal-I | 1.47 | 0.2250 | 1 | 1.000 | ns |
| agitated | positive | vocal-I | 0.47 | 0.4910 | 1 | 1.000 | ns |
| neutral | alert | vocal-II | 5.12 | 0.02360 | 1 | 0.1650 | ns |
| neutral | wary | vocal-II | 6.12 | 0.01330 | 1 | 0.1060 | ns |
| neutral | agitated | vocal-II | 9.96 | 0.00159 | 1 | 0.0159 | * |
| neutral | positive | vocal-II | 9.96 | 0.00159 | 1 | 0.0159 | * |
| alert | wary | vocal-II | 0.05 | 0.81900 | 1 | 1.0000 | ns |
| alert | agitated | vocal-II | 1.00 | 0.31700 | 1 | 1.0000 | ns |
| alert | positive | vocal-II | 1.00 | 0.31700 | 1 | 1.0000 | ns |
| wary | agitated | vocal-II | 0.60 | 0.43900 | 1 | 1.0000 | ns |
| wary | positive | vocal-II | 0.60 | 0.43900 | 1 | 1.0000 | ns |
| agitated | positive | vocal-II | 0.00 | 1.00000 | 1 | 1.0000 | ns |

S4.A.6. Glassmask-covered cue

| behaviour1 | behaviour2 | phase | statistic | p | df | p.adj | p.adj.signif |
|---|---|---|---|---|---|---|---|
| neutral | alert | silent | 15.24 | 9.45e-05 | 1 | 8.50e-04 | *** |
| neutral | wary | silent | 15.24 | 9.45e-05 | 1 | 8.50e-04 | *** |
| neutral | agitated | silent | 13.71 | 2.13e-04 | 1 | 1.49e-03 | ** |
| neutral | positive | silent | 30.11 | 0.00e+00 | 1 | 4.00e-07 | **** |
| alert | wary | silent | 0.00 | 1.00e+00 | 1 | 1.00e+00 | ns |
| alert | agitated | silent | 0.05 | 8.08e-01 | 1 | 1.00e+00 | ns |
| alert | positive | silent | 5.44 | 1.96e-02 | 1 | 9.80e-02 | ns |
| wary | agitated | silent | 0.05 | 8.08e-01 | 1 | 1.00e+00 | ns |
| wary | positive | silent | 5.44 | 1.96e-02 | 1 | 9.80e-02 | ns |
| agitated | positive | silent | 6.40 | 1.14e-02 | 1 | 6.84e-02 | ns |
| neutral | wary | vocal-I | 6.36 | 0.011600 | 1 | 0.09280 | ns |
| neutral | agitated | vocal-I | 0.33 | 0.564000 | 1 | 1.00000 | ns |
| neutral | positive | vocal-I | 6.36 | 0.011600 | 1 | 0.09280 | ns |
| neutral | alert | vocal-I | 1.68 | 0.194000 | 1 | 0.58200 | ns |
| alert | wary | vocal-I | 13.37 | 0.000256 | 1 | 0.00256 | ** |
| alert | agitated | vocal-I | 3.45 | 0.063000 | 1 | 0.27300 | ns |

| behaviour1 | behaviour2 | phase | statistic | p | df | p.adj | p.adj.signif |
|---|---|---|---|---|---|---|---|
| alert | positive | vocal-I | 13.37 | 0.000256 | 1 | 0.00256 | ** |
| wary | agitated | vocal-I | 4.00 | 0.045500 | 1 | 0.27300 | ns |
| wary | positive | vocal-I | 0.00 | 1.000000 | 1 | 1.00000 | ns |
| agitated | positive | vocal-I | 4.00 | 0.045500 | 1 | 0.27300 | ns |
| neutral | alert | vocal-II | 8.52 | 3.50e-03 | 1 | 0.024500 | * |
| neutral | wary | vocal-II | 12.60 | 3.86e-04 | 1 | 0.003090 | ** |
| neutral | agitated | vocal-II | 16.03 | 6.23e-05 | 1 | 0.000561 | *** |
| neutral | positive | vocal-II | 18.00 | 2.21e-05 | 1 | 0.000221 | *** |
| alert | wary | vocal-II | 0.52 | 4.67e-01 | 1 | 1.000000 | ns |
| alert | agitated | vocal-II | 1.66 | 1.97e-01 | 1 | 0.985000 | ns |
| alert | positive | vocal-II | 2.57 | 1.09e-01 | 1 | 0.654000 | ns |
| wary | agitated | vocal-II | 0.33 | 5.64e-01 | 1 | 1.000000 | ns |
| wary | positive | vocal-II | 0.81 | 3.66e-01 | 1 | 1.000000 | ns |
| agitated | positive | vocal-II | 0.11 | 7.39e-01 | 1 | 1.000000 | ns |

## B) Familiar

S4.B.1. Uncovered cue

| behaviour1 | behaviour2 | phase | statistic | p | df | p.adj | p.adj.signif |
|---|---|---|---|---|---|---|---|
| neutral | alert | silent | 4.50 | 3.39e-02 | 1 | 3.39e-01 | ns |
| neutral | anr | silent | 2.77 | 9.56e-02 | 1 | 6.69e-01 | ns |
| neutral | relaxed | silent | 4.50 | 3.39e-02 | 1 | 3.39e-01 | ns |
| neutral | affiliative | silent | 4.50 | 3.39e-02 | 1 | 3.39e-01 | ns |
| neutral | excited | silent | 15.15 | 9.89e-05 | 1 | 1.09e-03 | ** |
| alert | anr | silent | 0.33 | 5.64e-01 | 1 | 1.00e+00 | ns |
| alert | relaxed | silent | 0.00 | 1.00e+00 | 1 | 1.00e+00 | ns |
| alert | affiliative | silent | 0.00 | 1.00e+00 | 1 | 1.00e+00 | ns |
| alert | excited | silent | 28.12 | 1.00e-07 | 1 | 1.70e-06 | **** |
| anr | relaxed | silent | 0.33 | 5.64e-01 | 1 | 1.00e+00 | ns |
| anr | affiliative | silent | 0.33 | 5.64e-01 | 1 | 1.00e+00 | ns |
| anr | excited | silent | 25.48 | 4.00e-07 | 1 | 5.30e-06 | **** |
| relaxed | affiliative | silent | 0.00 | 1.00e+00 | 1 | 1.00e+00 | ns |
| relaxed | excited | silent | 28.12 | 1.00e-07 | 1 | 1.70e-06 | **** |
| affiliative | excited | silent | 28.12 | 1.00e-07 | 1 | 1.70e-06 | **** |
| alert | excited | vocal-I | 17.19 | 3.38e-05 | 1 | 3.38e-05 | **** |

S4.B.2. Hoodie- covered cue

| behaviour1 | behaviour2 | phase | statistic | p | df | p.adj | p.adj.signif |
|---|---|---|---|---|---|---|---|
| neutral | alert | silent | 5.44 | 1.96e-02 | 1 | 1.76e-01 | ns |
| neutral | anr | silent | 5.44 | 1.96e-02 | 1 | 1.76e-01 | ns |
| neutral | relaxed | silent | 2.27 | 1.32e-01 | 1 | 7.92e-01 | ns |
| neutral | affiliative | silent | 8.00 | 4.68e-03 | 1 | 4.68e-02 | * |
| neutral | excited | silent | 9.52 | 2.02e-03 | 1 | 2.22e-02 | * |
| alert | anr | silent | 0.00 | 1.00e+00 | 1 | 1.00e+00 | ns |
| alert | relaxed | silent | 1.00 | 3.17e-01 | 1 | 1.00e+00 | ns |
| alert | affiliative | silent | 1.00 | 3.17e-01 | 1 | 1.00e+00 | ns |
| alert | excited | silent | 23.14 | 1.50e-06 | 1 | 2.10e-05 | **** |
| anr | relaxed | silent | 1.00 | 3.17e-01 | 1 | 1.00e+00 | ns |
| anr | affiliative | silent | 1.00 | 3.17e-01 | 1 | 1.00e+00 | ns |
| anr | excited | silent | 23.14 | 1.50e-06 | 1 | 2.10e-05 | **** |
| relaxed | affiliative | silent | 3.00 | 8.33e-02 | 1 | 5.83e-01 | ns |
| relaxed | excited | silent | 18.24 | 1.95e-05 | 1 | 2.34e-04 | *** |
| affiliative | excited | silent | 26.00 | 3.00e-07 | 1 | 5.10e-06 | **** |
| neutral | alert | vocal-I | 1.00 | 0.317000 | 1 | 1.0000 | ns |

S4.B.2. Hoodie- covered cue

| behaviour1 | behaviour2 | phase | statistic | p | df | p.adj | p.adj.signif |
|---|---|---|---|---|---|---|---|
| neutral | anr | vocal-I | 0.00 | 1.000000 | 1 | 1.0000 | ns |
| neutral | relaxed | vocal-I | 0.00 | 1.000000 | 1 | 1.0000 | ns |
| neutral | affiliative | vocal-I | 1.00 | 0.317000 | 1 | 1.0000 | ns |
| neutral | excited | vocal-I | 8.33 | 0.003890 | 1 | 0.0545 | ns |
| alert | anr | vocal-I | 1.00 | 0.317000 | 1 | 1.0000 | ns |
| alert | relaxed | vocal-I | 1.00 | 0.317000 | 1 | 1.0000 | ns |
| alert | affiliative | vocal-I | 3.00 | 0.083300 | 1 | 0.8330 | ns |
| alert | excited | vocal-I | 4.57 | 0.032500 | 1 | 0.3580 | ns |
| anr | relaxed | vocal-I | 0.00 | 1.000000 | 1 | 1.0000 | ns |
| anr | affiliative | vocal-I | 1.00 | 0.317000 | 1 | 1.0000 | ns |
| anr | excited | vocal-I | 8.33 | 0.003890 | 1 | 0.0545 | ns |
| relaxed | affiliative | vocal-I | 1.00 | 0.317000 | 1 | 1.0000 | ns |
| relaxed | excited | vocal-I | 8.33 | 0.003890 | 1 | 0.0545 | ns |
| affiliative | excited | vocal-I | 11.00 | 0.000911 | 1 | 0.0137 | * |

S4.B.3. All-covered cue

| behaviour1 | behaviour2 | phase | statistic | p | df | p.adj | p.adj.signif |
|---|---|---|---|---|---|---|---|
| neutral | alert | silent | 1.14 | 2.85e-01 | 1 | 1.000000 | ns |
| neutral | anr | silent | 1.14 | 2.85e-01 | 1 | 1.000000 | ns |
| neutral | relaxed | silent | 9.00 | 2.70e-03 | 1 | 0.035100 | * |
| neutral | affiliative | silent | 6.40 | 1.14e-02 | 1 | 0.114000 | ns |
| neutral | excited | silent | 3.57 | 5.88e-02 | 1 | 0.412000 | ns |
| alert | anr | silent | 0.00 | 1.00e+00 | 1 | 1.000000 | ns |
| alert | relaxed | silent | 5.00 | 2.53e-02 | 1 | 0.228000 | ns |
| alert | affiliative | silent | 2.66 | 1.02e-01 | 1 | 0.612000 | ns |
| alert | excited | silent | 8.16 | 4.27e-03 | 1 | 0.051200 | ns |
| anr | relaxed | silent | 5.00 | 2.53e-02 | 1 | 0.228000 | ns |
| anr | affiliative | silent | 2.66 | 1.02e-01 | 1 | 0.612000 | ns |
| anr | excited | silent | 8.16 | 4.27e-03 | 1 | 0.051200 | ns |
| relaxed | affiliative | silent | 1.00 | 3.17e-01 | 1 | 1.000000 | ns |
| relaxed | excited | silent | 19.00 | 1.31e-05 | 1 | 0.000196 | *** |
| affiliative | excited | silent | 16.20 | 5.70e-05 | 1 | 0.000798 | *** |
| neutral | alert | vocal-I | 3.00 | 0.0833 | 1 | 0.750 | ns |

| behaviour1 | behaviour2 | phase | statistic | p | df | p.adj | p.adj.signif |
|---|---|---|---|---|---|---|---|

| behaviour1 | behaviour2 | phase | statistic | p | df | p.adj | p.adj.signif |
|---|---|---|---|---|---|---|---|
| neutral | anr | vocal-I | 6.40 | 0.0114 | 1 | 0.171 | ns |
| neutral | relaxed | vocal-I | 4.45 | 0.0348 | 1 | 0.383 | ns |
| neutral | affiliative | vocal-I | 6.40 | 0.0114 | 1 | 0.171 | ns |
| neutral | excited | vocal-I | 0.05 | 0.8080 | 1 | 1.000 | ns |
| alert | anr | vocal-I | 1.00 | 0.3170 | 1 | 1.000 | ns |
| alert | relaxed | vocal-I | 0.20 | 0.6550 | 1 | 1.000 | ns |
| alert | affiliative | vocal-I | 1.00 | 0.3170 | 1 | 1.000 | ns |
| alert | excited | vocal-I | 2.27 | 0.1320 | 1 | 1.000 | ns |
| anr | relaxed | vocal-I | 0.33 | 0.5640 | 1 | 1.000 | ns |
| anr | affiliative | vocal-I | 0.00 | 1.0000 | 1 | 1.000 | ns |
| anr | excited | vocal-I | 5.44 | 0.0196 | 1 | 0.255 | ns |
| relaxed | affiliative | vocal-I | 0.33 | 0.5640 | 1 | 1.000 | ns |
| relaxed | excited | vocal-I | 3.60 | 0.0578 | 1 | 0.578 | ns |
| affiliative | excited | vocal-I | 5.44 | 0.0196 | 1 | 0.255 | ns |

S4.B.4. Mask-covered cue

| behaviour1 | behaviour2 | phase | statistic | p | df | p.adj | p.adj.signif |
|---|---|---|---|---|---|---|---|
| neutral | alert | silent | 9.30 | 2.28e-03 | 1 | 2.51e-02 | * |
| neutral | anr | silent | 9.30 | 2.28e-03 | 1 | 2.51e-02 | * |
| neutral | relaxed | silent | 5.40 | 2.01e-02 | 1 | 1.61e-01 | ns |
| neutral | affiliative | silent | 9.30 | 2.28e-03 | 1 | 2.51e-02 | * |
| neutral | excited | silent | 4.00 | 4.55e-02 | 1 | 3.18e-01 | ns |
| alert | anr | silent | 0.00 | 1.00e+00 | 1 | 1.00e+00 | ns |
| alert | relaxed | silent | 1.00 | 3.17e-01 | 1 | 1.00e+00 | ns |
| alert | affiliative | silent | 0.00 | 1.00e+00 | 1 | 1.00e+00 | ns |
| alert | excited | silent | 21.16 | 4.20e-06 | 1 | 6.33e-05 | **** |
| anr | relaxed | silent | 1.00 | 3.17e-01 | 1 | 1.00e+00 | ns |
| anr | affiliative | silent | 0.00 | 1.00e+00 | 1 | 1.00e+00 | ns |
| anr | excited | silent | 21.16 | 4.20e-06 | 1 | 6.33e-05 | **** |
| relaxed | affiliative | silent | 1.00 | 3.17e-01 | 1 | 1.00e+00 | ns |
| relaxed | excited | silent | 16.33 | 5.31e-05 | 1 | 6.37e-04 | *** |
| affiliative | excited | silent | 21.16 | 4.20e-06 | 1 | 6.33e-05 | **** |
| neutral | alert | vocal-I | 1.80 | 0.180000 | 1 | 1.00000 | ns |

| behaviour1 | behaviour2 | phase | statistic | p | df | p.adj | p.adj.signif |
|---|---|---|---|---|---|---|---|
| neutral | anr | vocal-I | 1.80 | 0.180000 | 1 | 1.00000 | ns |
| neutral | relaxed | vocal-I | 1.80 | 0.180000 | 1 | 1.00000 | ns |
| neutral | affiliative | vocal-I | 1.80 | 0.180000 | 1 | 1.00000 | ns |
| neutral | excited | vocal-I | 6.36 | 0.011600 | 1 | 0.12800 | ns |
| alert | anr | vocal-I | 0.00 | 1.000000 | 1 | 1.00000 | ns |
| alert | relaxed | vocal-I | 0.00 | 1.000000 | 1 | 1.00000 | ns |
| alert | affiliative | vocal-I | 0.00 | 1.000000 | 1 | 1.00000 | ns |
| alert | excited | vocal-I | 12.25 | 0.000465 | 1 | 0.00698 | ** |
| anr | relaxed | vocal-I | 0.00 | 1.000000 | 1 | 1.00000 | ns |
| anr | affiliative | vocal-I | 0.00 | 1.000000 | 1 | 1.00000 | ns |
| anr | excited | vocal-I | 12.25 | 0.000465 | 1 | 0.00698 | ** |
| relaxed | affiliative | vocal-I | 0.00 | 1.000000 | 1 | 1.00000 | ns |
| relaxed | excited | vocal-I | 12.25 | 0.000465 | 1 | 0.00698 | ** |
| affiliative | excited | vocal-I | 12.25 | 0.000465 | 1 | 0.00698 | ** |

S4.B.5. Sunglass-covered cue

| behaviour1 | behaviour2 | phase | statistic | p | df | p.adj | p.adj.signif |
|---|---|---|---|---|---|---|---|
| neutral | alert | silent | 8.89 | 2.86e-03 | 1 | 0.025700 | * |
| neutral | anr | silent | 16.00 | 6.33e-05 | 1 | 0.000823 | *** |
| neutral | relaxed | silent | 8.89 | 2.86e-03 | 1 | 0.025700 | * |
| neutral | affiliative | silent | 13.23 | 2.75e-04 | 1 | 0.003300 | ** |
| neutral | excited | silent | 0.44 | 5.05e-01 | 1 | 1.000000 | ns |
| alert | anr | silent | 3.00 | 8.33e-02 | 1 | 0.583000 | ns |
| alert | relaxed | silent | 0.00 | 1.00e+00 | 1 | 1.000000 | ns |
| alert | affiliative | silent | 1.00 | 3.17e-01 | 1 | 1.000000 | ns |
| alert | excited | silent | 12.56 | 3.93e-04 | 1 | 0.004320 | ** |
| anr | relaxed | silent | 3.00 | 8.33e-02 | 1 | 0.583000 | ns |
| anr | affiliative | silent | 1.00 | 3.17e-01 | 1 | 1.000000 | ns |
| anr | excited | silent | 20.00 | 7.70e-06 | 1 | 0.000116 | *** |
| relaxed | affiliative | silent | 1.00 | 3.17e-01 | 1 | 1.000000 | ns |
| relaxed | excited | silent | 12.56 | 3.93e-04 | 1 | 0.004320 | ** |
| affiliative | excited | silent | 17.19 | 3.38e-05 | 1 | 0.000473 | *** |
| neutral | alert | vocal-I | 1.00 | 3.17e-01 | 1 | 1.000000 | ns |

| behaviour1 | behaviour2 | phase | statistic | p | df | p.adj | p.adj.signif |
|---|---|---|---|---|---|---|---|
| neutral | anr | vocal-I | 3.00 | 8.33e-02 | 1 | 0.833000 | ns |
| neutral | relaxed | vocal-I | 1.00 | 3.17e-01 | 1 | 1.000000 | ns |
| neutral | affiliative | vocal-I | 0.20 | 6.55e-01 | 1 | 1.000000 | ns |
| neutral | excited | vocal-I | 10.71 | 1.06e-03 | 1 | 0.011700 | * |
| alert | anr | vocal-I | 1.00 | 3.17e-01 | 1 | 1.000000 | ns |
| alert | relaxed | vocal-I | 0.00 | 1.00e+00 | 1 | 1.000000 | ns |
| alert | affiliative | vocal-I | 0.33 | 5.64e-01 | 1 | 1.000000 | ns |
| alert | excited | vocal-I | 15.21 | 9.62e-05 | 1 | 0.001350 | ** |
| anr | relaxed | vocal-I | 1.00 | 3.17e-01 | 1 | 1.000000 | ns |
| anr | affiliative | vocal-I | 2.00 | 1.57e-01 | 1 | 1.000000 | ns |
| anr | excited | vocal-I | 18.00 | 2.21e-05 | 1 | 0.000332 | *** |
| relaxed | affiliative | vocal-I | 0.33 | 5.64e-01 | 1 | 1.000000 | ns |
| relaxed | excited | vocal-I | 15.21 | 9.62e-05 | 1 | 0.001350 | ** |
| affiliative | excited | vocal-I | 12.80 | 3.47e-04 | 1 | 0.004160 | ** |

S4.B.6. Glassmask-covered cue

| behaviour1 | behaviour2 | phase | statistic | p | df | p.adj | p.adj.signif |
|---|---|---|---|---|---|---|---|
| neutral | alert | silent | 1.00 | 0.317000 | 1 | 1.000000 | ns |
| neutral | anr | silent | 3.76 | 0.052200 | 1 | 0.522000 | ns |
| neutral | relaxed | silent | 3.76 | 0.052200 | 1 | 0.522000 | ns |
| neutral | affiliative | silent | 7.36 | 0.006660 | 1 | 0.079900 | ns |
| neutral | excited | silent | 2.79 | 0.094700 | 1 | 0.663000 | ns |
| alert | anr | silent | 1.00 | 0.317000 | 1 | 1.000000 | ns |
| alert | relaxed | silent | 1.00 | 0.317000 | 1 | 1.000000 | ns |
| alert | affiliative | silent | 3.57 | 0.058800 | 1 | 0.522000 | ns |
| alert | excited | silent | 6.76 | 0.009320 | 1 | 0.103000 | ns |
| anr | relaxed | silent | 0.00 | 1.000000 | 1 | 1.000000 | ns |
| anr | affiliative | silent | 1.000000 | 0.317000 | 1 | 1.000000 | ns |
| anr | excited | silent | 11.636364 | 0.000647 | 1 | 0.009060 | ** |
| relaxed | affiliative | silent | 1.000000 | 0.317000 | 1 | 1.000000 | ns |
| relaxed | excited | silent | 11.636364 | 0.000647 | 1 | 0.009060 | ** |
| affiliative | excited | silent | 16.200000 | 0.000057 | 1 | 0.000855 | *** |
| neutral | alert | vocal-l | 0.1428571 | 0.705000 | 1 | 1.0000 | ns |

S4.B.6. Glassmask-covered cue

| behaviour1 | behaviour2 | phase | statistic | p | df | p.adj | p.adj.signif |
|---|---|---|---|---|---|---|---|
| neutral | anr | vocal-I | 0.1428571 | 0.705000 | 1 | 1.0000 | ns |
| neutral | relaxed | vocal-I | 0.1428571 | 0.705000 | 1 | 1.0000 | ns |
| neutral | affiliative | vocal-I | 1.8000000 | 0.180000 | 1 | 1.0000 | ns |
| neutral | excited | vocal-I | 5.5555556 | 0.018400 | 1 | 0.2020 | ns |
| alert | anr | vocal-I | 0.0000000 | 1.000000 | 1 | 1.0000 | ns |
| alert | relaxed | vocal-I | 0.0000000 | 1.000000 | 1 | 1.0000 | ns |
| alert | affiliative | vocal-I | 1.0000000 | 0.317000 | 1 | 1.0000 | ns |
| alert | excited | vocal-I | 7.1176471 | 0.007630 | 1 | 0.1070 | ns |
| anr | relaxed | vocal-I | 0.0000000 | 1.000000 | 1 | 1.0000 | ns |
| anr | affiliative | vocal-I | 1.0000000 | 0.317000 | 1 | 1.0000 | ns |
| anr | excited | vocal-I | 7.1176471 | 0.007630 | 1 | 0.1070 | ns |
| relaxed | affiliative | vocal-I | 1.0000000 | 0.317000 | 1 | 1.0000 | ns |
| relaxed | excited | vocal-I | 7.1176471 | 0.007630 | 1 | 0.1070 | ns |
| affiliative | excited | vocal-I | 11.2666667 | 0.000789 | 1 | 0.0118 | * |

**Q8. Do dogs show a difference in active negative behaviours depending on the familiarity of the experimenter?**

Table S5. Results of chi-square test to show that active negative behaviour are displayed by fewer dogs than what would occur by chance in silent phase

| group | frequency | statistic | p | df | p.signif |
|---|---|---|---|---|---|
| unfamiliar | 77/373 | 128.58 | 8.37e-30 | 1 | **** |
| familiar | 12/248 | 202.32 | 6.50e-46 | 1 | **** |

**Q9.** Does the display of active negative behaviours vary with the type of facial cue and the phase presented to unfamiliar dogs?

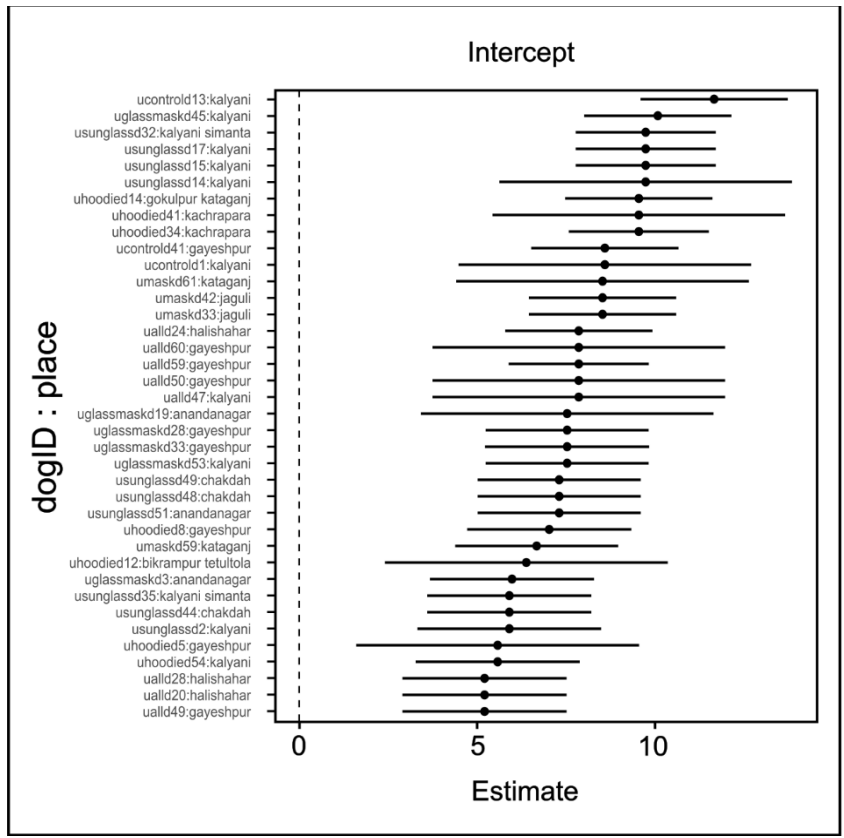

Figure S3: Random effect estimates for dogs whose 95% CI did not cross the 0 value

Table S6. Model-estimated marginal means of active negative behaviour probability of unfamiliar dogs across all cues and phases

| cue | phase | prob | se | asymp.LCL | Asymp.UCL |
|---|---|---|---|---|---|
| Uncovered | Silent | 0.00144 | 0.00234 | 5.96e-05 | 0.0346 |
| Uncovered | Vocal phase-I | 0.03006 | 0.03183 | 3.76e-03 | 0.2353 |
| Uncovered | Vocal phase-II | 0.03270 | 0.03557 | 3.86e-03 | 0.2701 |
| Hoodie | Silent | 0.02433 | 0.02911 | 2.32e-03 | 0.2487 |
| Hoodie | Vocal phase-I | 0.08063 | 0.08349 | 1.05e-02 | 0.5865 |
| Hoodie | Vocal phase-II | 0.06697 | 0.07422 | 7.55e-03 | 0.5615 |
| Sunglass | Silent | 0.02958 | 0.03229 | 3.47e-03 | 0.2466 |
| Sunglass | Vocal phase-I | 0.02984 | 0.03270 | 3.47e-03 | 0.2508 |
| Sunglass | Vocal phase-II | 0.06107 | 0.06270 | 8.10e-03 | 0.4418 |
| Mask | Silent | 0.00228 | 0.00307 | 1.63e-04 | 0.0319 |
| Mask | Vocal phase-I | 0.01660 | 0.01876 | 1.81e-03 | 0.1502 |
| Mask | Vocal phase-II | 0.05183 | 0.04927 | 7.99e-03 | 0.3266 |
| Sunglass + Mask | Silent | 0.07971 | 0.07897 | 1.13e-02 | 0.5342 |
| Sunglass + Mask | Vocal one | 0.03300 | 0.03766 | 3.50e-03 | 0.3016 |
| Sunglass + Mask | Vocal two | 0.01221 | 0.01561 | 9.95e-04 | 0.1476 |
| All covered | Silent | 0.18714 | 0.30682 | 7.14e-03 | 2.8500 |
| All covered | Vocal phase-I | 0.28437 | 0.45465 | 1.14e-02 | 3.2181 |
| All covered | Vocal phase-II | 0.59224 | 0.90300 | 2.52e-02 | 2.7233 |

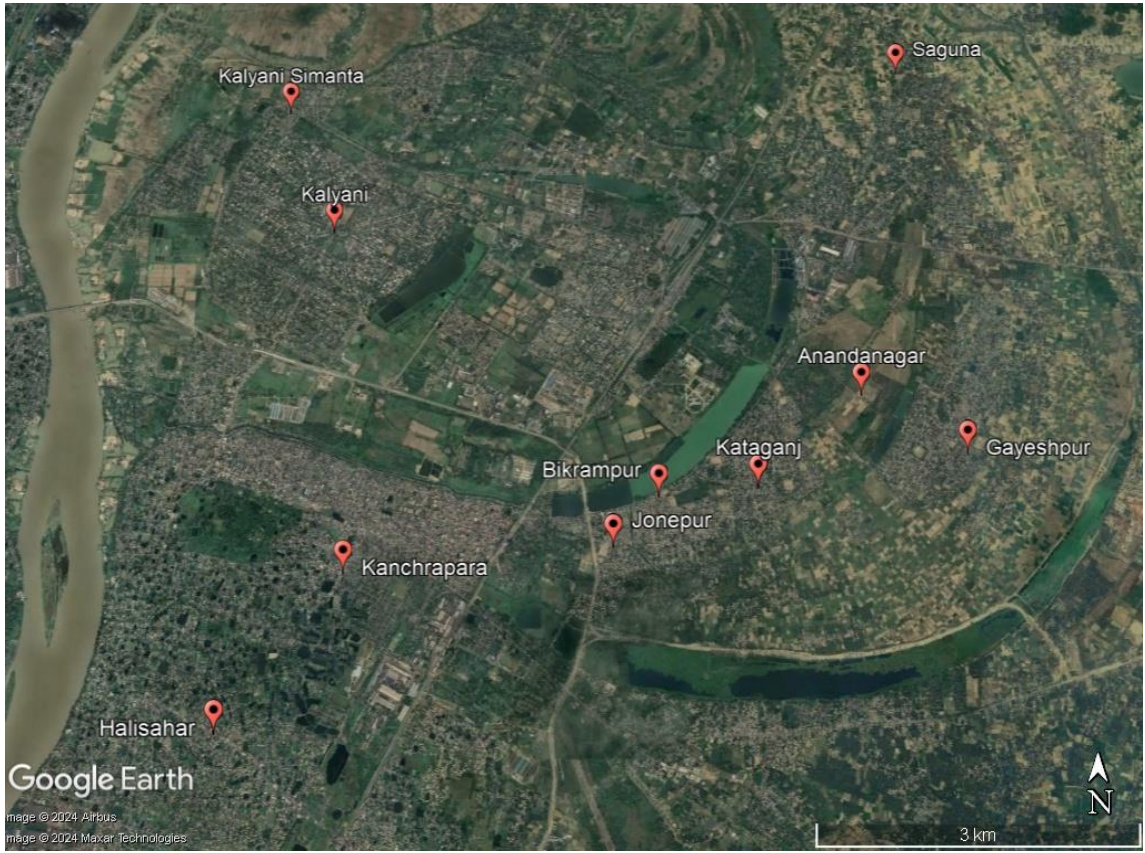
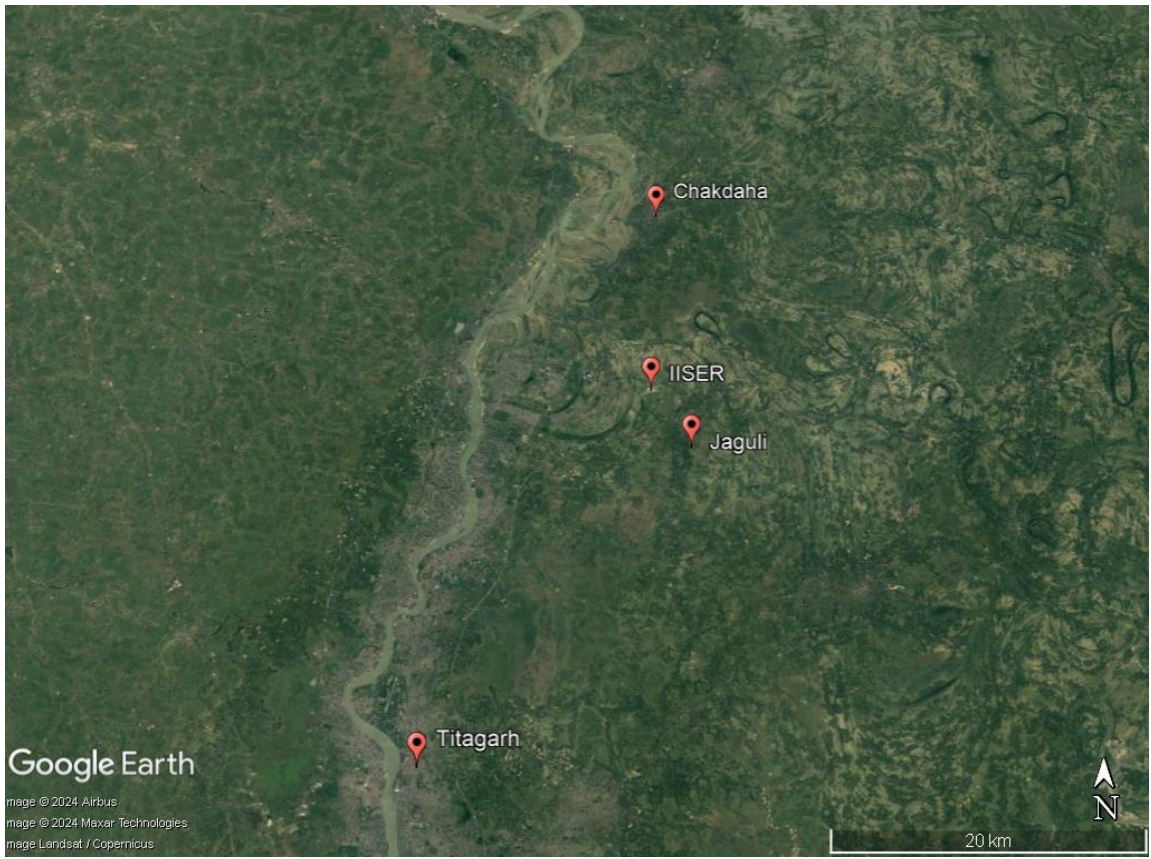

Figure S4: Map marking the field sites

## Behaviour and Analysis

Table S7: Behaviour ethogram along with scores

| Behaviour | Description | Direction | Value |
|---|---|---|---|
| Take notice | Dog looked at experimenter through open/half closed eyes on experimenter's /vocalisation/taking position or in the middle of the phase once | neutral | 0 |
| No attention | Dog did not pay attention to experimenter's approach/ vocalisation or closed eyes/looked away after taking notice | neutral | 0 |
| One-off gaze aversion | Looks back and forth and away from the experimenter once during a phase | neutral | 0 |
| Short stare | Continuously looking/gazing/staring at experimenter's head and neck region for <50% of the phase (<=3s for silent phase and <=5s for vocal one phase) | neutral | 0 |
| Head tilt | Dog tilts head on either left or right side as a tool to process stimuli | neutral | 0 |
| Frequent gaze aversion | Dog looks at experimenter and then looks away and repeats behaviour two or more times during a phase | negative | 1 |
| Continuous stare | Dog continuously looks/gazes/stares at experimenter's head and neck region for >50% of the phase (>3s for silent phase and >5s for vocal one phase) | negative | 1 |
| Pricked ears | Dog raises ears as a sign of alertness and vigilance on seeing experimenter approach or on hearing vocalisation | negative | 1 |
| Head raise | Dog lifts head or head and neck region as a sign of alertness and vigilance on | negative | 1 |

|  | seeing experimenter approach or on hearing vocalisation |  |  |
|---|---|---|---|
| Flinch | Dog makes a sudden, nervous movement in response to experimenter or his movement to face the dog | negative | 1 |
| Rigid posture | The dog tenses its body in response to the experimenter | negative | 1 |
| Peeled ears | Dogs flatten ears on the side of the head | positive | 1 |
| Recoil | Dog instinctively pulls back from the experimenter. There is a visible drawing back of the body and the dog shows rigid posture, half-sitting-half standing, ready to move away without actually moving away | negative | 2 |
| Sat up | A dog that was lying down or resting sits up quickly with rigid posture while maintaining continuous staring and/or does frequent gaze aversion and/or recoils and/or shows startled behaviour | negative | 2 |
| Tense stand-up | A dog that was sitting stands up while maintaining continuous staring and/or does frequent gaze aversion and/or recoils and/or shows startled behaviour | negative | 2 |
| Look back | While moving away, dog stops to check the status of experimenter. Involves movement of head and neck region | negative | 2 |
| Left tail wag | Dog wags tail only to the left. The direction of the tail base should be observed, if possible | negative | 2 |
| Pendulum tail wag | The tail stays down and is rapidly wagged from side to side like a pendulum | negative | 2 |
| Tucked tail | The tail droops and is tucked between legs and is a sign of submission and fear | negative | 2 |
| Relaxed stand up | The dog takes its time standing up like slowly picking itself up and/or doesn't look at experimenter while standing up and/or there is no pulling away motion and/or dog stands up to approach | positive | 2 |
| Right tail wag | Dog wags tail only to the right. The direction of the tail base should be observed, if possible | positive | 2 |
| Affiliative tail wag | Circular and/or loose back and forth wagging of tail vigorously, may also cause movement of hind part | positive | 2 |
| Bark | Make a loud, short sound when threatened or for showing aggression | negative | 3 |
| Growl | Low frequency rolling sound for showing aggression | negative | 3 |
| Purr | Low, soft happy sound for showing affiliation ("kuin-kuin" | positive | 3 |
| Move away | Dog walks/moves/runs away/keeps moving | negative | 4 |
| Hide | Dog walks behind an object (tarpaulin/board/wall) and peeks from behind it | negative | 4 |
| Moves forward | Dog starts walking towards experimenter without actually approaching | positive | 4 |
| Belly display | Dog lies down on back and shows belly to experimenter | positive | 4 |
| Approach | Dog approaches experimenter within petting distance | positive | 5 |